\documentclass[
preprint,
showpacs, showkeys,
nofootinbib,
 amsmath,amssymb,
 aps,
 pra,
floatfix,
]{revtex4-1}

\pdfoutput=1
\usepackage{graphicx}

\usepackage{dcolumn}
\usepackage{bm}

\usepackage{subfigure}
\usepackage{longtable}
\usepackage{lipsum}
\usepackage{color}
\usepackage{comment}

\newcommand{\HeatFlux}{\boldsymbol{\mathcal{Q}}}
\newcommand{\SpeciesFlux}{\boldsymbol{\mathcal{F}}}

\newcommand{\StressTensor}{\boldsymbol{\Pi}}
\newcommand{\ShearViscosity}{\eta}
\newcommand{\BulkViscosity}{\kappa}

\newcommand{\EntropyProduction}{\mathfrak{v}}

\newcommand{\ChemicalSpeciesLabel}{\mathfrak{M}}
\newcommand{\StochioPM}{\nu}

\newcommand{\U}{\mathbf{U}}
\newcommand{\FH}{\mathbf{F}_H}
\newcommand{\FD}{\mathbf{F}_D}
\newcommand{\FS}{\mathbf{F}_S}
\newcommand{\HH}{\mathbf{H}}

\newcommand{\half}{\frac{1}{2}}

\global\long\def\V#1{\boldsymbol{#1}}
\global\long\def\M#1{\boldsymbol{#1}}

\global\long\def\D#1{\Delta#1}
\global\long\def\d#1{\delta#1}

\global\long\def\grad{\M{\nabla}}

\global\long\def\av#1{\left\langle #1\right\rangle }

\newcommand{\ZSpeciesFlux}{\M{\mathcal{Z}}^{\SpeciesFlux}}
\newcommand{\ZHeatFlux}{\M{\mathcal{Z}}^{\HeatFlux}}
\newcommand{\ZOmega}{\mathcal{Z}^\Omega}
\newcommand{\ZSpeciesFluxC}{{\mathcal{Z}}^{\SpeciesFlux}}
\newcommand{\ZHeatFluxC}{{\mathcal{Z}}^{\HeatFlux}}

\newcommand{\modified}[1]{#1}

\begin{document}

\title{Fluctuating hydrodynamics of multi-species reactive mixtures}

\author{Amit Kumar Bhattacharjee$^{1}$,  Kaushik Balakrishnan$^{2}$, Alejandro L. Garcia$^{3}$, John B. Bell$^{4}$ and Aleksandar Donev$^{1}$}
\affiliation{$^1$ Courant Institute of Mathematical Sciences, New York University \\
 251 Mercer Street, New York, NY 10012 \\ }
\affiliation{$^2$ Jet Propulsion Laboratory, 4800 Oak Grove Drive, Pasadena CA 91109 \\}
\affiliation{$^3$ Department of Physics and Astronomy, San Jose State University \\
 1 Washington Square, San Jose, CA 95192 \\ }
\affiliation{$^4$ Computational Research Division, Lawrence Berkeley National Laboratory \\
 1 Cyclotron Road, Berkeley, CA 94720 \\ }

\date{\today}

\begin{abstract}
We formulate and study computationally the fluctuating compressible Navier-Stokes equations for reactive multi-species fluid mixtures. We contrast two different expressions for the covariance of the stochastic chemical production rate in the Langevin formulation of stochastic chemistry, and compare both of them to predictions of the chemical Master Equation for homogeneous well-mixed systems close to and far from thermodynamic equilibrium. We develop a numerical scheme for inhomogeneous reactive flows, based on our previous methods for non-reactive mixtures [K. Balakrishnan, A. L. Garcia, A. Donev and J. B. Bell, Phys. Rev. E 89:013017, 2014]. We study the suppression of non-equilibrium long-ranged correlations of concentration fluctuations by chemical reactions, as well as the enhancement of pattern formation by spontaneous fluctuations. Good agreement with available theory demonstrates that the  formulation is robust and a useful tool in the study of fluctuations in reactive multi-species fluids. At the same time, several problems with Langevin formulations of stochastic chemistry are identified, suggesting that future work should examine combining Langevin and Master Equation descriptions of hydrodynamic and chemical fluctuations.
\end{abstract}


\maketitle

\section{Introduction}
\label{sec:intro}

Chemical reactions are of central importance in both natural and industrial processes spanning the range of length scales from the microscopic, through the mesoscopic, and up to macroscopic scales. It is the rule, rather than the exception, that chemical reactions are strongly coupled to hydrodynamic transport processes, such as advection, diffusion, and thermal conduction. Prominent examples include diffusion-limited aggregation, pattern and chemical wave formation in reactive solutions, reaction-driven convective instabilities, heterogeneous catalysis, combustion, complex biological processes, and others. \modified{Even in a homogeneous system with only slightly exothermic reactions, the chemistry is coupled to the hydrodynamics, leading to non trivial effects such as fluctuation induced transitions \cite{Detonation_Fluctuations}}.

Fluctuations affect reactive systems in multiple ways. In stochastic biochemical systems, such as reactions inside the cytoplasm, or in catalytic processes, some of the reacting molecules are present in very small numbers and therefore discrete stochastic models are necessary to describe the system. In diffusion-limited reactive systems, such as simple coagulation \modified{$2A \rightarrow A_2$} or annihilation $A+B \rightarrow C$, spatial fluctuations in the concentration of the reactants grow as the reaction progresses and must be accounted for to accurately model the correct macroscopic behavior.~\cite{Annihilation_AplusB,Coagulation_Renormalization}
In unstable systems, such as diffusion-driven Turing instabilities \cite{Turing_Fluctuations1,Turing_Fluctuations2,Turing_RDME,Turing_RDME_2,Turing_MD}, detonation \cite{Detonation_Fluctuations}, or buoyancy-driven convective instabilities \cite{InstabilityRT_Chemistry}, fluctuations are responsible for initiating
the instability and may profoundly affect its subsequent temporal and spatial development.
In systems with a marginally-stable manifold, fluctuations lead to a drift along this manifold that cannot be described by the traditional law of mass action, and has been suggested as being an important mechanism in the emergence of life \cite{MarginalStability_Brogioli,MarginalStability_Brogioli2,MarginalStability_LMA}.

Much of the work on modeling stochastic chemistry has been for homogeneous, ``well-mixed'' systems,
such as continuously stirred tank reactors (CSTRs), but there is
increasing interest in spatial models~\cite{StochChemSpatialNature}.
When hydrodynamic transport is included, the focus has almost exclusively been on species diffusion, and there is a large body of literature on stochastic reaction-diffusion models.
A Master Equation approach, notably, the Chemical Master Equation (CME), is widely accepted for modelling well-mixed systems.
The Reaction-Diffusion Master Equation (RDME) extends this type of approach to spatially-varying systems \cite{MalekRDME1975,nicolis1977self,GardinerBook}. In the RDME, the system is subdivided into reactive subvolumes (cells) and diffusion is modeled as a discrete random walk by particles moving between cells, while reactions are modeled using local CMEs \cite{RDME_Review_Erban,MesoRD}. A large number of efficient and elaborate event-driven kinetic Monte Carlo algorithms for solving the CME and RDME, exactly or approximately, have been developed with many tracing their origins to the Stochastic Simulation Algorithm (SSA) of Gillespie~\cite{Gillespie76,GillespieSSAreview}.
The issue of convergence as the
RDME grid is refined is delicate.~\cite{MesoRD_GridResolution,RDME_Bimolecular_Petzold,CRDME}
Variants of the RDME have been proposed that improve or eliminate the sensitivity of
the results to the grid resolution, such as the convergent RDME (CRDME) of Isaacson \cite{CRDME} in which reactions can happen between molecules in neighboring cells as well.
Particle-based spatial methods for stochastic chemistry include reactive Brownian dynamics~\cite{ReactiveBrownianDynamics,Smoldyn},
Green's Function Reaction Dynamics~\cite{GFRD_PRL,GFRD_KMC},
first-passage kinetic Monte Carlo~\cite{FPKMC1,FPKMC2,FPKMC_drift,FPKMC_lattice},
the small-voxel tracking algorithm~\cite{VoxelTracking_Gillespie}, and others.

In large part, the development of stochastic reaction-diffusion models has been divorced from
work in the fluid dynamics community.
Full hydrodynamic transport including advection, sound waves, viscous stress,
thermal conduction, etc., as well as nonequilibrium thermodynamics and chemistry,
are fairly common in the reacting flow community. See, for example, textbooks by Kuo \cite{Kuo_combustion}
and Law \cite{Law_combustion}.
However work in this area focuses on macroscopic modeling; spontaneous thermal fluctuations, either chemical or hydrodynamic, are typically not considered.

Within the field of nonequilibrium thermodynamics, the fluctuation-dissipation theorem provides
the connection between hydrodynamic transport and spontaneous fluctuations.
In particular, as an extension of conventional hydrodynamic theory, fluctuating hydrodynamics
incorporates mesoscopic fluctuations in a fluid by adding stochastic flux terms
to the deterministic fluid equations \cite{FluctHydroNonEq_Book}.
These noise terms are white in space and time and are
formulated using fluctuation-dissipation relations
to yield equilibrium covariances of the fluctuations in agreement with equilibrium statistical mechanics.
Linearized fluctuating hydrodynamics was first introduced by Landau and Lifshitz \cite{Landau:Fluid} and
has since been used to study simple and binary fluid systems in and out of equilibrium \cite{FluctHydroNonEq_Book}.

A number of numerical algorithms for solving the equations of fluctuating hydrodynamics have also been developed
\cite{Garcia:87b,LB_ThermalFluctuations,Bell_10,LLNS_S_k,StaggeredFluct_Energy,MultispeciesCompressible,LowMachExplicit,LowMachMultispecies}.
These algorithms draw from a wealth of deterministic computational fluid dynamics (CFD) techniques
and handle transport such as diffusion
in a much more sophisticated fashion than random hopping between cells.
For example, they include effects such as cross-diffusion, barodiffusion, thermodiffusion (i.e., Soret effect)
as well as advection by fluid motion. Furthermore, semi-implicit temporal discretizations
and higher-order spatial discretizations can be used, even as fluctuations due to the discrete nature
of the fluid are accounted for. In this spirit, Koh and Blackwell~\cite{HoppingSSA_Gradient}
propose using a more traditional gradient-driven diffusive flux formulation consistent with CFD practice
within an CME-based description;
however, their treatment of fluctuations is rather \textit{ad hoc} and
not consistent with the formulation of stochastic mass fluxes in fluctuating hydrodynamics.
Very recently a spatial chemical Langevin formulation (SCLE) was proposed~\cite{Spatial_CLE}
in which the chemical Langevin approximation~\cite{ChemicalLangevin_Gillespie}
is applied to the RDME treating diffusive hops as another reaction in a very large reaction network.
While this leads to a formulation similar to fluctuating hydrodynamics it has several shortcomings, notably,
it does not allow one to treat diffusion using advanced CFD algorithms.

Investigations utilizing fluctuating hydrodynamics have revealed the crucial
importance of hydrodynamic fluctuations in transport mechanisms, especially mass and heat diffusion.
Notably, it is now well-known that all nonequilibrium diffusive mixing processes are accompanied
by long-range correlations of fluctuations. In certain scenarios these nonequilibrium fluctuations
grow in physical extent well beyond molecular scales with magnitudes far greater than those of equilibrium fluctuations.
These so-called ``giant fluctuations'' are observed in laboratory experiments
\cite{GiantFluctConcentration_Sengers,GiantFluctuations_Nature,FractalDiffusion_Microgravity}, and arise
because of the coupling between thermal velocity fluctuations and concentration or temperature
fluctuations. In fact, it has recently been shown using nonlinear fluctuating hydrodynamics
that mass diffusion in liquids is dominated by advection by thermal velocity fluctuations \cite{DiffusionJSTAT}.
Therefore, modeling diffusion using collections of independent random walkers,
as done in the RDME, is fundamentally inappropriate
for describing the nature of hydrodynamic fluctuations at microscopic and mesoscopic scales; instead,
hydrodynamic coupling (correlations) between the diffusing particles must be taken into account \cite{DDFT_Hydro}.
Including fluctuations within the continuum description has also been shown to be important
in particle-continuum hybrids \cite{DSMC_Hybrid},
and should also benefit hybrid models for reaction-diffusion systems \cite{BD_PDE_Hybrid}.

In the hydrodynamic equations chemical reactions may be treated as
a white noise source term~\cite{haken2004synergetics,GardinerBook},
in a fashion analogous to the stochastic transport fluxes.
The study of fluctuating hydrodynamic models that include chemical reactions is relatively recent
and there are few computational studies in the literature.
Stochastic reaction-diffusion equations are considered by Atzberger in \cite{AMR_ReactionDiffusion_Atzberger},
but only within the reaction-diffusion framework and not accounting for fluctuations in the chemical production rates.
A thorough discussion of stochastic formulations of chemical reactions
within the framework of statistical mechanics can be found in the monograph by Keizer \cite{KeizerBook};
Keizer does not, however, consider hydrodynamic transport in spatially-extended systems in depth.
In a sequence of important papers \cite{FluctChemistry_Rubi,FluctuatingReactionDiffusion,LLNS_ReactionDiffusion,LLNS_ReactionDiffusion2}, chemical reactions have been incorporated in a {\em nonlinear} nonequilibrium thermodynamic formalism, making it possible to combine realistic nonlinear deterministic models based on the traditional law of mass action (LMA) kinetics with fluctuating hydrodynamics. When considering fluctuations, however, a linearized approximation was used by the authors, limiting the range of applicability to modeling small Gaussian fluctuations around a macroscopic state that evolves in a manner unaffected by the fluctuations. A more phenomenological approach was followed to fit the LMA into the nonequilibrium thermodynamics GENERIC formalism by Grmela and Ottinger~\cite{GENERIC_Chemical}, but fluctuations were not considered. Here, we formulate a complete set of fluctuating hydrodynamic equations for a reactive multispecies mixture of ideal gases. We account for mass, momentum and energy transport, and chemical reactions, and consider a nonlinear formalism for describing the thermal fluctuations.

Hydrodynamics is a macroscopic coarse-grained description, and fluctuating hydrodynamics is a mesoscopic coarse-grained description.
As such, both descriptions rely on the approximation that the length and time scales under consideration are much larger than
molecular, i.e, that each coarse-grained degree of freedom involves an average over many molecules.
In fact, although formally written as a continuum model, fluctuating hydrodynamics is, in truth, a discrete
model that only makes sense when seen as a coarse-grained description for the
evolution of a collection of spatially-discrete hydrodynamic variables involving averages over many nearby molecules \cite{DiscreteLLNS_Espanol,FluctDiff_FEM}.
The fact that many molecules are involved in the reactions allows for a Langevin-like
continuum description (i.e., diffusion processes) of the fluctuations
instead of discrete models such as master equations (i.e., jump processes).
The accuracy of Langevin formulations for chemically reacting systems has long been
a topic of debate \cite{CLE_vs_vanKampen,ChemistryReview_Gillespie}.
In this work, we take the first step in combining realistic fluid dynamics with a stochastic chemical description
and adopt a Langevin approach to describing fluctuations. In future work, we will explore combining Langevin and ME approaches
together, thus further bridging the apparent gap between the two.

Here, we first formulate the fluctuating reactive Navier-Stokes-Fourier equations,
discuss their physical validity, and develop numerical methods for solving the stochastic partial differential equations.
The methodology is a direct extension of our previous work on fluctuating hydrodynamics
for non-reactive multispecies gas mixtures \cite{MultispeciesCompressible}
to include a Langevin model of chemical reactions.
We consider two distinct Langevin models, which are identical when very close
to chemical equilibrium but differ far from thermodynamic equilibrium.
The first model, which we term the Log-Mean equation (LME), is based on the GENERIC formulation of Grmela and Ottinger~\cite{GENERIC_Chemical}, but can be traced to older work on the subject as well \cite{ChemicalFluctuations_FPE,BistableChemical_FPE,KeizerBook}.
The second Langevin model is the more familiar Chemical Langevin Equation \cite{ChemicalLangevin_Gillespie,KeizerBook,KurtzTheorem_CLE}.

The resulting algorithms are used to assess the importance of thermal fluctuations
in several simple but relevant examples.
The first example is a simple dimerization reaction, which has been studied theoretically in prior work by others \cite{FluctuatingReactionDiffusion,LLNS_ReactionDiffusion,LLNS_ReactionDiffusion2}.
Our second example is the Baras-Pearson-Mansour (BPM) reaction network~\cite{Baras1990,Baras1996},
which exhibits a rich behavior ranging from bistability to limit cycles.
We study these examples in both well-mixed small-scale systems,
comparing with the Chemical Master Equation,
and in spatially-extended systems, comparing with fluctuating hydrodynamic theory and previous numerical work.
For the latter, rather than imposing the non-equilibrium constraint by fixing concentrations in the bulk,
the constraints are applied as boundary conditions,
thus maintaining \emph{strict} consistency with equilibrium thermodynamics,
including microscopic reversibility (detailed balance), in all of the models we study.
These examples illustrate how thermal fluctuations drive giant concentration fluctuations
and how they affect the rate of pattern formation in an inhomogeneous system.



\section{Theory}
\label{sec:form}

In this section, we summarize the mathematical formulation of the complete
fluctuating Navier-Stokes (FNS) equations for compressible reactive multispecies fluid mixtures.
The details for non-reactive fluid mixtures are presented in \cite{MultispeciesCompressible};
here we focus on the chemistry.
The formulation is first presented in its general form;
the specific case of reactions in ideal gas mixtures is treated in Section~\ref{LMAsection}.

The species density, momentum and energy equations of hydrodynamics for a mixture of $N_\mathrm{S}$ species are given by
\begin{equation}
\frac{\partial }{\partial t} \left( \rho_s \right)  + {\bf \nabla} \cdot \left( \rho_s {\bf v} \right) +
{\bf \nabla} \cdot {\SpeciesFlux}_s =  m_s \Omega_s , \qquad (s=1,\ldots N_\mathrm{S})
\label{eq:spec}
\end{equation}
\begin{equation}
\frac{\partial }{\partial t} \left( \rho {\bf v} \right)
+ {\bf \nabla} \cdot \left[ \rho {\bf v} {\bf v}^T +  p \mathbf{I} \right]
+ {\bf \nabla} \cdot  {\StressTensor}  = \rho {\bf g},
\label{eq:mom}
\end{equation}
\begin{equation}
\frac{\partial }{\partial t} \left( \rho E \right)  + {\bf \nabla} \cdot \left[ (\rho E + p) {\bf v}  \right] +
{\bf \nabla} \cdot \left[ {\HeatFlux} + \StressTensor \cdot {\bf v} \right] = \rho {\bf v \cdot g},
\label{eq:energy}
\end{equation}
where $\rho_s$, $m_s$, and $\Omega_s$ are the mass density, molecular mass,
and number density production rate for species $s$.
The variables ${\bf v}$, $p$, and $E$ denote, respectively,
fluid velocity, pressure, and specific total energy for the mixture.
The total density is $\rho = \sum_{s=1}^{N_\mathrm{S}} \rho_s$,
${\bf g}$ is gravitational acceleration, and superscript $T$ denotes transpose.

We consider
a system with $N_\mathrm{R}$ elementary reactions with reaction $r$ written in the form,
\begin{equation*}
\mathfrak{R}_r : \qquad
\sum_{s=1}^{N_\mathrm{S}} \StochioPM_{sr}^+ \ChemicalSpeciesLabel_s \rightleftarrows
\sum_{s=1}^{N_\mathrm{S}} \StochioPM_{sr}^- \ChemicalSpeciesLabel_s
\end{equation*}
Here $\ChemicalSpeciesLabel_s$ are the chemical symbols and $\StochioPM_{si}^\pm$ are the
molecule numbers for the forward and reverse reaction $r$.
The stoichiometric coefficients are $\nu_{sr}=\StochioPM_{sr}^{-}-\StochioPM_{sr}^{+}$
and mass conservation requires that $\sum_s \nu_{sr} m_r=0$.~\cite{KeizerBook}
For simplicity of notation, when there is no ambiguity we omit the range of the sums
and write $\sum_s$ for sums over all species,
and write $\sum_r$ for sums over all reactions.
Note that chemistry does not appear explicitly in the energy equation (\ref{eq:energy})
since the species heat of formation is included in the specific total energy.

Transport properties are given in terms of
the species diffusion flux, ${\SpeciesFlux}$, viscous tensor, ${\StressTensor}$, and heat flux, ${\HeatFlux}$.
Mass conservation requires that the species diffusion flux
and the production rate due to chemical reactions satisfy the constraints,
\begin{equation}
\sum_s \SpeciesFlux_s = 0
\qquad\mathrm{and}\qquad
\sum_s m_s \Omega_s = 0
\label{eq:det_sum_cond}
\end{equation}
so that summing the species equations gives the continuity equation,
\begin{equation}
\frac{\partial }{\partial t}  \rho   + {\bf \nabla} \cdot \left( \rho {\bf v} \right) = 0.
\label{eq:cont}
\end{equation}
The detailed form of the transport terms is summarized in Appendix~A, see \cite{MultispeciesCompressible} for details.
It is important to note that we neglect any possible effect of
the chemical reactions on the transport coefficients of the mixture.

We write the chemical production rate as the sum of a deterministic and a
stochastic contribution, $\Omega_s = \overline{\Omega}_s + \widetilde{\Omega}_s$,
with the stochastic rate going to zero in the deterministic
limit.~\cite{RD_DeterLimit_Arnold1980,RDME_Limits_Arnold}
To formulate these production rates
we define the dimensionless chemical affinity as,
\begin{equation}
\mathcal{A}_r
= - \frac{1}{k_B T} \sum_{s}\nu_{sr}m_{s}\mu_{s}
= \sum_{s}\StochioPM_{sr}^{+}\hat{\mu}_{s}
- \sum_{s}\StochioPM_{sr}^{-}\hat{\mu}_{s},
\label{AffinityEqn}
\end{equation}
where $\mu_s$ is the specific chemical potential (i.e., per unit mass) and
$\hat{\mu}_{s} = m_{s}\mu_{s}/k_B T$ is the dimensionless chemical potential per particle;
$T$ and $k_B$ are temperature and Boltzmann's constant, respectively.
Summing over reactions gives the deterministic production rate for species $s$ as~\cite{GENERIC_Chemical}
\begin{equation}
\overline{\Omega}_s = \sum_r \nu_{sr} \frac{p}{\tau_r k_B T} \hat{\mathcal{A}}_r
\label{eq:OmegaDet}
\end{equation}
where
\begin{equation}
\hat{\mathcal{A}}_r
=\exp\left(\sum_{s}\StochioPM_{sr}^{+}\hat{\mu}_{s}\right)
-\exp\left(\sum_{s}\StochioPM_{sr}^{-}\hat{\mu}_{s}\right)
=\prod_{s}e^{\StochioPM_{sr}^{+}\hat{\mu}_{s}}
-\prod_{s}e^{\StochioPM_{sr}^{-}\hat{\mu}_{s}}
\label{eq:AtildeGeneral}
\end{equation}
and $\tau_r$ is a time scale characterizing the reaction rate.
This form of the deterministic equations, while
at first sight appearing different from the more familiar law of mass action (LMA), is fully consistent with it. The
production rate, as given by (\ref{eq:OmegaDet}) and (\ref{eq:AtildeGeneral}), is also consistent with nonequilibrium thermodynamics \cite{FluctuatingReactionDiffusion};
this way of expressing the production rate in terms of a thermodynamic driving force
(difference of exponentials of chemical potentials) can be seen
as a generalization of the LMA to non-ideal systems,
as elaborated in Section~\ref{LMAsection}.

For a binary mixture undergoing a dimerization reaction, the deterministic part of the complete set of hydrodynamic equations including chemical reactions has been fit into
a {\em nonlinear} nonequilibrium thermodynamics formalism in Ref. \cite{FluctuatingReactionDiffusion} by introducing an additional reaction coordinate, as inspired by earlier work of Pagonabarraga et al. \cite{FluctChemistry_Rubi}.
This extends earlier considerations of dimerization reactions in a strictly linear fluctuating chemistry formalism \cite{FluctChemistry_Grossmann}.
In the limit of high reaction barrier the equations written in \cite{FluctuatingReactionDiffusion} are equivalent to the ones we employ here even though our notation is different.
However, fluctuating contributions in \cite{FluctuatingReactionDiffusion} are only considered in a linearized approximation, severely limiting the range of applicability to describing small Gaussian fluctuations around a deterministic average flow.

In next two sub-sections we develop two {\em nonlinear} forms for the stochastic contribution to the reactive production rates, one coming from irreversible thermodynamics cast in the GENERIC formalism \cite{GENERIC_Chemical},
and the other being a generalization of the more familiar form associated with the chemical Langevin equation (CLE) \cite{ChemicalLangevin_Gillespie,KeizerBook,KurtzTheorem_CLE}.

\subsection{The Log-Mean Equation}

Grmela and Ottinger~\cite{GENERIC_Chemical} cast the phenomenological LMA (\ref{eq:OmegaDet}) in the GENERIC formalism
and obtain a nonlinear form for the dissipative matrix, under the {\em assumption} of a quadratic dissipative potential.
Note that the entropy production rate can {\em uniquely} be written as a quadratic function of the thermodynamic driving force {\em only} for a single reaction; the resulting peculiar form of the mobility (dissipative) matrix (see Eq. (113) in \cite{GENERIC_Chemical}) involving a logarithmic mean has recently been justified from a model reduction perspective \cite{ModelReduction_GENERIC}.
Here we \emph{assume} that there is no cross-coupling between different reactions and thus associate an independent stochastic production rate with each reaction. Coupling between distinct reactions has been considered within a nonequilibrium thermodynamic framework only in some very specific cases \cite{ElectrochemicalCells_GENERIC,CoupledReactions_Bedeaux} and a general formulation requires more detailed knowledge about the coupling mechanism than is available in practice.

Following the general principles for including fluctuations
\footnote{Fluctuations are not considered by Grmela and Ottinger \cite{GENERIC_Chemical}, however, the ``square root'' of the mobility matrix written in Eq. (113) in \cite{GENERIC_Chemical} is straightforward to write, and this leads to the log-mean equation considered here.}
in the GENERIC formalism \cite{OttingerBook}, it is straightforward to write a Gaussian stochastic production rate assuming independence among the different reaction channels,
\begin{equation}
\widetilde{\Omega}^\mathrm{LM}_s = \sum_r \nu_{sr} \sqrt{2 \mathcal{D}^\mathrm{LM}_{r}}
\diamond \ZOmega_r
\label{eq:LME_Omega}
\end{equation}
where
\begin{equation}
\mathcal{D}_{r}^\mathrm{LM} =
\frac{p}{\tau_r k_B T}
\frac{\hat{\mathcal{A}}_r}{\mathcal{A}_r}
\label{eq:LME_Amplitude}
\end{equation}
and where $\ZOmega_r$ are independent white-noise random scalar fields with covariance
\[
\langle \ZOmega_{r}(\mathbf{r},t)
\ZOmega_{r^\prime}(\mathbf{r}',t') \rangle
= \delta_{r,r^\prime} \, \delta(\mathbf{r}-\mathbf{r}')\, \delta(t - t'),
\]
with each $\ZOmega_r$ driving the stochastic production rate of a single chemical reaction $r$.
We refer to this formulation for the stochastic chemistry as the
``log-mean'' form; the reasoning behind this name will become evident
when presented in Section~\ref{LMAsection} for ideal mixtures.
Note that (\ref{eq:LME_Omega}) uses the kinetic or Klimontovich interpretation~\cite{KlimontovichCalculus,KineticStochasticIntegral_Ottinger}
of the stochastic integral, formally denoted as a kinetic stochastic product with a $\diamond$ symbol in (\ref{eq:LME_Omega}).
The variance of the
stochastic forcing $\mathcal{D}_{r}^\mathrm{LM}$ can be seen
to be positive because $\hat{\mathcal{A}}$ and $\mathcal{A}$ always
have the same sign~\cite{GENERIC_Chemical}.
Note that $\sum_s m_s \overline{\Omega}_s =
\sum_s m_s \widetilde{\Omega}^\mathrm{LM}_s = 0$, as required by mass conservation.

For the purposes of exposition it is useful to consider
a homogeneous ``well-mixed'' system of volume $\D{V}$, which will correspond
to a single hydrodynamic cell after spatial discretization of (\ref{eq:spec}).
The dynamics of the (intensive) number density $n_s=N_s / \D{V}$, where $N_s=\D{V} \rho_s / m_s$
is the number of molecules of species $s$ in the cell, is given by,
\begin{equation}
\frac{d n_s(t)}{dt} =  \sum_r \nu_{sr} \left[
   \frac{p}{\tau_r k_B T} \hat{\mathcal{A}}_r
   + \sqrt{\frac{2 \mathcal{D}^\mathrm{LM}_{r}}{\D{V}}} \diamond \mathcal{W}^{\Omega}_r
   \right],
\label{eq:LME_sinetic}
\end{equation}
which can also be written in Ito form as,
\begin{equation}
\frac{d n_s(t)}{dt}  =  \sum_r \nu_{sr} \left[
   \frac{p}{\tau_r k_B T} \hat{\mathcal{A}}_r
   + \sqrt{\frac{2 \mathcal{D}^\mathrm{LM}_{r}}{\D{V}}} \, \mathcal{W}^{\Omega}_r
   + \frac{1}{\D{V}} \sum_{s'=1}^{N_\mathrm{S}} \nu_{s'r}\left(\frac{\partial \mathcal{D}_r^\mathrm{LM}}{\partial{n_{s'}}}\right)
   \right],
   \label{eq:LME}
\end{equation}
where $\mathcal{W}^{\Omega}_r(t)$ are independent scalar white noise processes with covariance
$\langle \mathcal{W}^{\Omega}_r(t) \mathcal{W}^{\Omega}_{r^\prime}(t^\prime) \rangle = \delta_{r,r^\prime} \delta(t-t^\prime)$.
We call this  system of stochastic ordinary differential equations (SODEs) the ``log-mean'' equation (LME).

The derivative in the last stochastic drift term in (\ref{eq:LME}) is the directional derivative of $\mathcal{D}_r^\mathrm{LM}$ along the reaction coordinate. Unlike the more familiar Ito or Stratonovich interpretations of the noise, the kinetic form of the noise ensures that the corresponding Fokker-Planck equation has the traditional form \cite{GrabertBook},
\[
\frac{\partial P(\V{n},t)}{\partial t} = \sum_r \sum_s\frac{\partial }{\partial n_s}
\left\{ \nu_{sr} \mathcal{D}_r^\mathrm{LM} \left[
- \mathcal{A}_r P
+ \frac{1}{\D{V}} \sum_{s'} \nu_{s'r} \frac{\partial P}{\partial n_{s'}}
\right] \right\},
\label{FPE_LME}
\]
This ensures that the LME is in detailed balance with respect to the Einstein distribution $\sim S / (k_B T)$ for a closed system at thermodynamic equilibrium,
where $S$ is the total entropy of the system \cite{OttingerBook}.
We note that it is not possible to obtain the LME from the chemical master equation (CME)
with a systematic procedure; one must invoke
some guiding principles about the structure of coarse-grained Fokker-Planck equations to ``derive'' this form of the noise \cite{ChemicalFluctuations_FPE,BistableChemical_FPE,GrabertBook}.

\subsection{The Chemical Langevin Equation}

Since both $\mathcal{A}$ and $\hat{\mathcal{A}}$ are equal to zero
at chemical equilibrium, near chemical equilibrium we can linearize (\ref{eq:LME_Amplitude})
to first order in the affinity $\mathcal{A}$,
and approximate the amplitude of the stochastic production rate
in terms of a sum over each forward and reverse reaction, that is,
\begin{equation}
\frac{\hat{\mathcal{A}}_r}{\mathcal{A}_r}\approx
\exp\left(\sum_{s}\StochioPM_{sr}^{+}\hat{\mu}_{s}\right)
+\exp\left(\sum_{s}\StochioPM_{sr}^{-}\hat{\mu}_{s}\right)
= \prod_{s}e^{\StochioPM_{sr}^{+}\hat{\mu}_{s}}
+ \prod_{s}e^{\StochioPM_{sr}^{-}\hat{\mu}_{s}}.
\label{eq:linearized}
\end{equation}
Since this sum of products of exponentials is evidently positive, we can potentially use it even far from chemical equilibrium, and write the stochastic production rate as,
\begin{equation}
\widetilde{\Omega}_s^\mathrm{CL}
= \sum_r \nu_{sr} \left( \sqrt{2 \mathcal{D}_{r,+}^\mathrm{CL}}\, \ZOmega_{r,+}
                                  +\sqrt{2 \mathcal{D}_{r,-}^\mathrm{CL}}\, \ZOmega_{r,-} \right)
\label{eq:CLE_Omega}
\end{equation}
where \footnote{Note that the two terms in (\ref{eq:CLE_Omega}) can be combined together to give a single white-noise process with amplitude $\mathcal{D}_{r}^\mathrm{CL} = \mathcal{D}_{r,+}^\mathrm{CL} + \mathcal{D}_{r,-}^\mathrm{CL}$; this leaves the covariance of the stochastic forcing unchanged.}
\begin{equation}
\mathcal{D}_{r,\pm}^\mathrm{CL} = \half \left( \frac{p}{\tau_r k_B T} \right)
\prod_{s}e^{\StochioPM_{sr}^{\pm}\hat{\mu}_{s}}.
\label{eq:CLE_Amplitude}
\end{equation}
Here $\ZOmega_{r,+}$ are independent white-noise scalar random fields that give the stochastic contribution from the forward reaction, while $\ZOmega_{r,-}$ correspond to the reverse reactions; the forward and reverse reactions are taken to be independent.
In the next section the production rate factors, $\mathcal{D}^\mathrm{LM}_{r}$ and $\mathcal{D}_{r}^\mathrm{CL}$,
are further simplified for the case of ideal gas mixtures.

The form (\ref{eq:CLE_Omega}) for the amplitude of the stochastic production rate is found
in most work on the subject \cite{ChemicalLangevin_Gillespie,NonEqThermo_Bedeaux,FluctuatingReactionDiffusion,LLNS_ReactionDiffusion,LLNS_ReactionDiffusion2}.
For example, though not written in this form, eqn.~(8f) in Ref. \cite{LLNS_ReactionDiffusion2}
contains a sum of two exponential terms and is equivalent to (\ref{eq:linearized}) for the specific reaction considered there
\footnote{As the authors note in a later publication \cite{LLNS_ReactionDiffusion2}, the sign in Eq. (88) in Ref. \cite{FluctuatingReactionDiffusion} is wrong and should be a plus rather than a minus. With this change that equation is equivalent to (\ref{eq:CLE_Omega}) evaluated at the deterministic solution, in the spirit of linearized fluctuating hydrodynamics.}.
For a well-mixed homogeneous system of volume $\D{V}$, the number densities of molecules of the different species follow the system of SODEs,
\begin{equation}
\frac{d n_s(t)}{dt}  =  \sum_r  \nu_{sr} \left[
   \frac{p}{\tau_r k_B T} \hat{\mathcal{A}}_r
   + \sqrt{\frac{2 \mathcal{D}^\mathrm{CL}_{r}}{\D{V}}} \mathcal{W}^{\Omega}_r
   \right] \;\;\; .
\label{eq:CLE}
\end{equation}
The stochastic equation (\ref{eq:CLE}) is commonly referred to as the
chemical Langevin equation (CLE) following Gillespie~\cite{ChemicalLangevin_Gillespie}, and can be obtained from the CME by an uncontrolled truncation of the Kramers-Moyal expansion to second order. It is traditional to
assume an Ito interpretation of the noise in the CLE, even though no precise justification for this can be made within
the accuracy to which the CLE approximates the CME \cite{KurtzTheorem_CLE}. Mathematically, the nonlinear CLE contains similar information to the central limit theorem (i.e., linearized fluctuating hydrodynamics) corresponding to the CME in the limit of weak noise (large number of reactant molecules).

As seen from (\ref{eq:linearized}), the two stochastic differential equations for the number densities,
the LME using the kinetic noise (\ref{eq:LME})
and the CLE using the Ito noise (\ref{eq:CLE}),
are equivalent near chemical equilibrium.
They are, however, different far from chemical equilibrium, as we illustrate in more detail in Section~\ref{WellMixedResultsSection}. Notably, the forward and reverse reactions are treated together in the LME, consistent with the fact that, due to microscopic reversibility, there is only one independent rate coefficient for each reaction.~\cite{KeizerBook}
The ratio of the forward and reverse reaction rates is related to the equilibrium reaction constant, which is a {\em thermodynamic} and not a kinetic quantity.
In fact, the LME is closely-related to the notion of the existence of a state of thermodynamic equilibrium in which each pair of forward and reverse reactions are in detailed balance with each other; one cannot write an LME for a system with irreversible reactions, which fundamentally violate detailed balance.

By contrast, the forward and reverse reactions are treated completely independently in the CLE and there is no difficulty in writing a CLE for a system with irreversible reactions. The CLE is evidently inconsistent with the notion of detailed balance and is, in fact, inconsistent with equilibrium thermodynamics. Although written in a different form, Keizer's (4.8.37) is the CLE, and Keizer's (4.8.36) is the LME \cite{KeizerBook}; Keizer argues that the CLE is the correct equation and concludes: ``Although the theoretical description of nonequilibrium ensembles would be greatly simplified if the phenomenological choice [LME] were correct, this appears not to be the case.'' We will compare and contrast these two equations on some specific examples in Section~\ref{WellMixedResultsSection}.

\subsection{The Law of Mass Action and Ideal Gas Mixtures}\label{LMAsection}

In the formulation of hydrodynamic transport one normally works with the specific chemical potential,
which has the general form,~\cite{IrrevThermoBook_Kuiken}
\[
\mu_s(p,T,\V{X}) = \frac{k_B T}{m_s} \left(\ln X_s +\ln \gamma_s\right)  + \mu_s^o(p,T),
\]
where $\mu_s^o$ is the chemical potential at a reference state,
$X_s=N_s / \sum_{s^\prime} N_{s^\prime}$ is the mole fraction, and
$\gamma_s(p,T,\V{X})$ is the activity coefficient of species $s$.
For chemistry it is more convenient to work with
a dimensionless chemical potential per particle,
\[
\hat{\mu}_s = \frac{m_s \mu_s}{k_B T}
= \ln (X_s \gamma_s) + \hat{\mu}_s^o,
\]
where $\hat{\mu}_s^o = (m_s \mu_s^o)/(k_B T)$.
Note that $X_s \gamma_s$ is the activity (i.e., effective concentration) and for an ideal mixture, $\gamma_s=1$.~\cite{PrigogineKondepudi_ThermoTextbook}
This gives
\[
\exp\left(\StochioPM_{sr}^{\pm}\hat{\mu}_{s}\right) =
\exp \left(\StochioPM_{sr}^\pm \hat{\mu}_s^o\right) \left( X_s \gamma_s \right)^{\StochioPM_{sr}^\pm} \;\; ,
\]
which leads to a generalized law-of-mass action (LMA) of the form
\begin{eqnarray}
\overline{\Omega}_s = \sum_r
\nu_{sr} \frac{p}{\tau_r k_B T}
\hat{\mathcal{A}}_r
&=& \sum_r
\nu_{sr} \left( \kappa_r^{+} \prod_{s^\prime} \left( X_{s^\prime} \gamma_{s^\prime} \right)^{\StochioPM_{{s^\prime}r}^+}
- \kappa_r^{-} \prod_{s^\prime} \left( X_{s^\prime} \gamma_{s^\prime} \right)^{\StochioPM_{{s^\prime}r}^-}\right),
\label{eq:LMA_generalized}
\end{eqnarray}
where $\kappa_r^{\pm}(T,p,\V{X})$ are the more familiar forward/reverse
reaction rates (per unit time and per unit volume).
Since there is only one independent timescale parameter, $\tau_r$, the forward and reverse rates
are not independent and the LMA gives the ratio to be the equilibrium constant,
\begin{equation}
K_r(p,T)=\frac{\kappa_r^{+}}{\kappa_r^{-}}
= \left[ \frac{\prod_{s}\left( X_{s} \gamma_{s} \right)^{\StochioPM_{{k}r}^-}}
{\prod_{s}\left( X_{s} \gamma_{s} \right)^{\StochioPM_{{k}r}^+}}  \right]_\mathrm{eq}
=\exp\left(-\sum_s \nu_{sr} \hat{\mu}_s^o \right)
,
\label{eq:eq_const}
\end{equation}
which is a purely thermodynamic quantity (closely related to the dimensionless reference Gibbs energy for the reaction at
a unit reference pressure) that can be calculated from pure component data~\cite{zemansky1981heat,NonEqThermo_Bedeaux}.
Note that in chemistry texts the equilibrium constant is typically defined
in terms of concentrations rather than activities as we have done here.

For ideal gas mixtures we can further simplify the generalized LMA (\ref{eq:LMA_generalized})
to the more familiar form using number densities instead of mole fractions.
From classical statistical mechanics, for an ideal gas mixture we can write\footnote{This
is in the classical regime, that is, when the molecules' mean spacing is much larger than their de Broglie wavelength.}
\begin{equation}
\hat{\mu}_s = \ln n_{s} + \ln \left(\frac{\Lambda_s^3(T)}{j_s(T)}\right),
\label{eqn:mu_rdealgas}
\end{equation}
where $\Lambda_s = h/\sqrt{2\pi m_s k_B T}$ is the thermal wavelength of a structure-less particle and
$j(T)$ is the partition function for the internal degrees of freedom.~\cite{pathria2011statistical}
In general $j(T)$ is a complicated function depending on the quantized energy levels of a molecule
but in the classical approximation $j(T) = (T/T_o)^{\frac12 z}$ where $z$ is the number of classical
internal degrees of freedom and $T_o$ is a reference temperature.

For ideal gas mixtures the chemical production rate (\ref{eq:LMA_generalized}) can be written
in the familiar power-law form,
\begin{equation}
\overline{\Omega}_s
= \sum_r \nu_{sr} \left( k_r^{+} \prod_{s^\prime} n_{s^\prime}^{\StochioPM_{s^\prime r}^+}
- k_r^{-} \prod_{s^\prime} n_{s^\prime}^{\StochioPM_{s^\prime r}^-}\right),
\label{eq:LMA_numdens}
\end{equation}
where $k_r^{\pm}$ are the forward/reverse reaction rates for the LMA formulated in
terms of number density instead of activity.
For uni-molecular reactions (e.g., $\mathfrak{M}_1 \rightarrow \ldots$) the ``decay time'' for a particle
is usually assumed to be constant, in which case
the corresponding reaction rate (e.g., $k_r^+$) is a constant.
For bi-molecular reactions (e.g., $\mathfrak{M}_1 + \mathfrak{M}_2 \rightarrow \ldots$)
the production rate is usually assumed to be proportional to the collision frequency
times an Arrhenius factor, in which case the corresponding reaction rate
is only a function of temperature.~\cite{MulticomponentBook_Giovangigli}


For the stochastic production rate in an ideal gas mixture, using,
\[
\mathcal{A}_r
=\sum_{s}\StochioPM_{sr}^{+}\hat{\mu}_{s}
-\sum_{s}\StochioPM_{sr}^{-}\hat{\mu}_{s}
=\ln\frac
{\exp\left(\sum_{s}\StochioPM_{sr}^{+}\hat{\mu}_{s}\right)}
{\exp\left(\sum_{s}\StochioPM_{sr}^{-}\hat{\mu}_{s}\right)}
=\ln\frac
{k_r^{+}\prod_{s} n_{s}^{\StochioPM_{sr}^{+}}}
{k_r^{-}\prod_{s} n_{s}^{\StochioPM_{sr}^{-}}}.
\]
gives
\begin{eqnarray}
\mathcal{D}_{r}^\mathrm{LM} &=&
\frac{k_r^{+}\prod_{s} n_{s}^{\StochioPM_{sr}^{+}}
-k_r^{-}\prod_{s} n_{s}^{\StochioPM_{sr}^{-}}}
{\ln\left(k_r^{+}\prod_{s} n_{s}^{\StochioPM_{sr}^{+}}\right)
-\ln\left(k_r^{-}\prod_{s} n_{s}^{\StochioPM_{sr}^{-}}\right)}
= \mathrm{logmean}\left\{
k_r^{+}\prod_{s} n_{s}^{\StochioPM_{sr}^{+}},\;
k_r^{-}\prod_{s} n_{s}^{\StochioPM_{sr}^{-}}
\right\}
\label{eq:LME_Amp_Ideal}
\\
\mathcal{D}_{r}^\mathrm{CL} &=& \half \left[
k_r^{+}\prod_{s} n_{s}^{\StochioPM_{sr}^{+}}
+k_r^{-}\prod_{s} n_{s}^{\StochioPM_{sr}^{-}}
\right]
= \mathrm{arthmean}\left\{
k_r^{+}\prod_{s} n_{s}^{\StochioPM_{sr}^{+}},\;
k_r^{-}\prod_{s} n_{s}^{\StochioPM_{sr}^{-}}
\right\}.
\label{eq:CLE_Amp_Ideal}
\end{eqnarray}
where $\mathrm{logmean}$ and $\mathrm{arthmean}$ are the logarithmic and arithmetic mean, respectively.
Note that $\mathcal{D}_{r}^\mathrm{LM}$ is zero if either reaction rate equals zero
while $\mathcal{D}_{r}^\mathrm{CL}$ is non-zero if either reaction rate is non-zero.
The logarithmic mean form of the noise in the ideal mixture case has appeared, phenomenologically, in several early papers \cite{ChemicalFluctuations_FPE,BistableChemical_FPE} and in
more recent work \cite{ModelReduction_GENERIC,LogarithmicMean_Mielke,LogarithmicMean_Mielke2} by other authors.


\section{Numerical Scheme}
\label{sec:numerical}

The numerical integration of (\ref{eq:spec})-(\ref{eq:energy})
is based on a method of lines approach in which
we discretize the equations in space and then use an SODE integration algorithm
to advance the solution using the basic overall approach described in ~\cite{MultispeciesCompressible}.
The spatial discretization uses a finite volume representation with cell volume
$\D{V}$, where $\U_{i,j,k}^n$ denotes the average value of
$\U = (\rho_s, \rho \mathbf{v}, \rho E)$ in cell-$(i,j,k)$ at time step $n$.
To ensure that the algorithm satisfies discrete fluctuation-dissipation
balance, the spatial discretizations for the hydrodynamic fluxes are done using centered
discretizations; see~\cite{LLNS_S_k} and \cite{MultispeciesCompressible} for details.

Discretization of the system in space results in a system of SODEs
%
driven by a collection of independent white-noise processes $\V{\mathcal{W}}(t)$ that represent a spatial discretization of
the random Gaussian fields $\V{\mathcal{Z}}(\V{r},t)$ used to construct the noise. After temporal discretization
these white noise processes are represented by a collection of i.i.d. standard normal variates $\V{Z}$,
which can be thought of as a spatio-temporal discretization of $\V{\mathcal{Z}}$;
the discretization is reflected in the presence of a prefactor $1/\sqrt{\D{V} \Delta t}$
in the expressions for $\widetilde{\Omega}$ given below \cite{DFDB}.
%

For temporal integration we use the low-storage third-order Runge-Kutta (RK3) scheme previously
used to solve the single and two-component FNS equations \cite{LLNS_S_k}, using the
weighting of the stochastic forcing proposed by Delong {\it et al.} \cite{DFDB}.
With this choice of weights, the temporal
integration is weakly second-order accurate for additive noise (e.g., the linearized
equations of fluctuating hydrodynamics \cite{MultiscaleIntegrators}).
As discussed at length in Ref. \cite{MultiscaleIntegrators} the hydrodynamic stochastic
fluxes should be considered as additive noise in a linearized approximation.

The implementation of the methodology
supports three boundary conditions in addition to periodicity.
The first is a specular, adiabatic wall which is impermeable to momentum, mass or heat transport (i.e., all fluxes are zero at the wall).
A second type of boundary condition is a no slip, reservoir wall at which the normal velocity vanishes
(i.e., the total mass flux at the boundary vanishes)
and the other velocity components, mole fractions and temperature satisfy inhomogeneous
Dirichlet boundary conditions; this mimics a permeable membrane connected to a reservoir on the other side of the boundary.
The third boundary condition is a variant of the no slip condition
for which the wall is impermeable to mass but conducts heat.
When a Dirichlet condition is specified for a given quantity, the corresponding diffusive
flux is computed as a difference of the cell-center value and the value on the boundary.
In such cases the corresponding stochastic flux is multiplied by $\sqrt{2}$ to ensure
discrete fluctuation-dissipation balance, as explained in detail \cite{LLNS_Staggered,LowMachExplicit}.

Because the noise arising from the chemical reactions is multiplicative
special care must be taken to capture the correct stochastic drift terms 
arising from the kinetic interpretation of the noise in the \modified{LME.~\cite{kloeden2010numerical,ottinger2012stochastic}}
We have chosen to write the equations in Ito form, which leads to an additional stochastic drift term in the LME (\ref{eq:LME}).
To integrate the Ito form in time, we evaluate the amplitude of the noise at the beginning of the time step and reuse the same random increments in all three stages of the RK3 scheme. The stochastic drift term arising in the LME is treated as a deterministic term but is also only evaluated at the beginning of the time step. The resulting scheme is only first-order weakly accurate. It is possible to construct second-order weak integrators by using the special one-dimensional nature (i.e., there is only a single reaction coordinate for each reaction even if there are many species involved) of the chemical noise \cite{WeakTrapezoidal}. However, in our simulations the time step is typically limited by stability considerations for advective and diffusive hydrodynamic processes, and therefore chemistry is accurately resolved even by a first-order scheme. Alternative temporal integration strategies will be discussed in the Conclusions.

The chemical Langevin form of the noise in (\ref{eq:CLE}) is discretized as,
\begin{equation}
(\widetilde{\Omega}^\mathrm{CL}_s)_{i,j,k}^n
= \sum_{r} \nu_{sr}
\sqrt{\frac{2(\mathcal{D}^\mathrm{CL}_{r})_{i,j,k}^n}{\D{V} \Delta t}}\,
{(Z_r^\Omega)_{i,j,k}^n},
\qquad\qquad s=1,\ldots,N_\mathrm{S},
\label{eq:CLE_Omega_Discrete}
\end{equation}
where $\left( Z_r^\Omega \right)_{i,j,k}$ are zero-mean normal Gaussian variates generated independently in each cell at the beginning of each time step,
$\langle (Z_r^\Omega)_{i,j,k}^n (Z_{r'}^\Omega)_{i',j',k'}^{n'}\rangle = \delta_{i,i'}\delta_{j,j'}\delta_{k,k'}\delta_{n,n'}\delta_{r,r'}.$
Note that the two terms in (\ref{eq:CLE_Omega}) have been combined into a single white-noise process with amplitude $\mathcal{D}_{r}^\mathrm{CL} = \mathcal{D}_{r,+}^\mathrm{CL} + \mathcal{D}_{r,-}^\mathrm{CL}$.
As discussed above, the LME noise (\ref{eq:LME_Omega}) leads to an Ito correction
in (\ref{eq:LME}),
\[
(\widetilde{\Omega}_s^\mathrm{LM})_{i,j,k}^n
= \sum_{r=1}^{N_\mathrm{R}} \nu_{sr} \left[
\sqrt{\frac{2(\mathcal{D}_{r}^\mathrm{LM})_{i,j,k}^n}{\D{V} \Delta t}}\,
{(Z_r^\Omega)_{i,j,k}^n}
+ \frac{1}{\D{V}}
\sum_{s'=1}^{N_\mathrm{S}} \nu_{s'r}\left(\frac{\partial \mathcal{D}_r^\mathrm{LM}}{\partial{n_{s'}}}\right)
\right].
\label{LME_discretized}
\]
The directional derivative of $\mathcal{D}_r^\mathrm{LM}$ in the last term can be evaluated analytically, or, for simplicity of implementation, it can be efficiently approximated numerically using a finite difference along the reaction coordinate.

\section{Numerical Results}
\label{sec:results}

In this section we describe several test problems
that demonstrate the capabilities of the numerical methodology.
We consider two reaction systems, the first being simple dimerization,
\modified{
\begin{equation}
\mathfrak{R}_1: \qquad
2A \leftrightarrows A_2
\label{DimerizationEqn}
\end{equation}
}
where $\mathfrak{M} = (A,A_2)$, that is, species~1 is the monomer and species~2 is the dimer.
In Section~\ref{WellMixedDimerizationSection} we investigate simple dimerization in a homogeneous system;
in Section~\ref{GiantFluctuationsSection} we investigate the ``giant fluctuation''
phenomenon in the presence of dimerization for a system with an applied concentration gradient.

The second model we consider is based on the
Gray-Scott (GS) model~\cite{GrayScott1,GrayScott2}, which is known to
exhibit a rich morphology of stationary
and time-dependent patterns~\cite{Pearson09071993}.
This model, as formulated by Pearson~\cite{Pearson09071993}, consists of the reactions,
\begin{eqnarray}
&\mathfrak{R}_1:& \qquad
U + 2V \rightarrow 3V \nonumber \\
&\mathfrak{R}_2:& \qquad
V \rightarrow S \label{GSmodelEqn}\\
&\mathfrak{R}_3:& \qquad
U \leftrightarrows U_f \nonumber \\
&\mathfrak{R}_4:& \qquad
V \rightarrow V_f \nonumber
\end{eqnarray}
where the concentrations of the ``feed species'', $U_f$ and $V_f$, are held fixed
and species $S$ is inert.
Since elementary reactions are rarely trimolecular in nature,
we consider a variation of the GS model developed by Baras \textit{et al.}~\cite{Baras1990,Baras1996}.
The Baras-Pearson-Mansour (BPM) model\footnote{Also known as the OLR model.} is,
\modified{
\begin{eqnarray}
&\mathfrak{R}_1:& \qquad
U + W \leftrightarrows V + W \nonumber \\
&\mathfrak{R}_2:& \qquad
2V \leftrightarrows W + S \nonumber \\
&\mathfrak{R}_3:& \qquad
V \leftrightarrows S \label{BPMmodelEqn}\\
&\mathfrak{R}_4:& \qquad
U \leftrightarrows U_f \nonumber \\
&\mathfrak{R}_5:& \qquad
V \leftrightarrows V_f \nonumber
\end{eqnarray}
}
with $\mathfrak{M} = (U,V,W,S,U_f,V_f)$.
This model was developed as a more realistic variant of the Gray-Scott model
suitable for particle simulations of dilute
gases.\footnote{To implement the BPM model
in a molecular simulation with binary collisions Baras \textit{et al.}
modify the model by making all the reactions bimolecular
and by introducing an ``auxiliary'' species, $A$.}
Note that the first and third reactions
are irreversible in the original BPM model, which is not consistent with detailed balance.
The BPM model has not been studied as extensively as the GS model
but its dynamics are expected to be qualitatively similar.
Note that the BPM model replaces the trimolecular reaction in the GS model with
a pair of bimolecular reactions and introduces $W$ as an intermediary species.
In the standard BPM model \cite{Baras1990} the number densities of
$S$, $U_f$, and $V_f$ are held fixed so, being an open system, total mass
is not conserved and detailed balance is not satisfied.~\cite{KeizerBook}
In Section~\ref{WellMixedBpmSection} we investigate this standard BPM model
in a homogeneous ``well-mixed'' system;
in Section~\ref{PatternFormationSection} we simulate a two dimensional
domain with full hydrodynamic transport with
species $S$, $U_f$, and $V_f$ held fixed only at the boundaries.


\subsection{Homogeneous Systems}\label{WellMixedResultsSection}

We first consider homogeneous ``well-mixed''
\footnote{The precise mathematical definition of a ``well-mixed'' system is delicate and will not
be discussed here at length. Roughly speaking, it means that diffusion is sufficiently fast
compared to reactions; see \cite{RDME_Limits_Arnold} for a theorem on the limit of infinite diffusion.}
systems of volume $\D{V}$ with only chemistry (i.e., no hydrodynamics).
In this section we compare the results obtained using the log-mean equation (LME) form,
(\ref{eq:LME_Amp_Ideal}), and the chemical Langevin equation (CLE) form, (\ref{eq:CLE_Amp_Ideal}),
with results from CME simulations performed using the Stochastic Simulation Algorithm (SSA),
also known as the Gillespie algorithm.~\cite{Gillespie76}
The chemical master equation (CME) is widely accepted as
an accurate model for well-mixed chemical systems
and SSA is a popular scheme for simulating the
stochastic process described by the CME.~\cite{GillespieSSAreview}
As we will see, the two forms for the Langevin noise have their advantages and disadvantages
and both forms are only approximations of the CME with limited ranges of validity.

\subsubsection{Dimerization Reaction}\label{WellMixedDimerizationSection}

We start by considering the dimerization reaction (\ref{DimerizationEqn})
in a closed system for which
the deterministic production rate for species 1 (monomers) is
\[
\overline{\Omega}_1 = - 2 (k^{+} n_1^2 - k^{-} n_2),
\]
and by mass conservation, for dimers $\Omega_2 = - {\textstyle \frac12} \Omega_1$ since $m_2 = 2 m_1$.
The constraint of mass conservation can also be expressed as
$
n_{1} + 2 n_2 = n_{0},
$
where $n_0$ is the initial number density of A particles (in either monomer or dimer form). We may then write,
\[
\overline{\Omega}_1 = - 2 k^{+} n_1^2 + k^{-} (n_0 - n_1)
\]
and limit our attention to the monomer species.
For simplicity we take the ratio of the rate constants to be
$
k^{+}/k^{-}= 1/n_{0}
$
so the equilibrium mass fraction $(Y_1)_\mathrm{eq} = \frac12$
(i.e., $\overline{\Omega}_1=0$ for $Y_1= \frac12$).

The log-mean stochastic production rate for $n_1$ is
\[
\widetilde{\Omega}_1 = -2 \sqrt{\frac{2 \mathcal{D}^\mathrm{LM}}{\D{V}}}\diamond \mathcal{W}^\Omega
\qquad\mathrm{with}\qquad
\mathcal{D}^\mathrm{LM} =
\frac{k^{+}n_1^{2}-k^{-}n_2}{\ln\left(k^{+}n_1^{2}\right)-\ln\left(k^{-}n_2\right)}.
\]
From the corresponding Fokker-Planck equation (FPE) (\ref{FPE_LME}) one finds the
LME is in detailed balance with respect to the equilibrium distribution
\begin{equation}
P^\mathrm{LM}_{\text{eq}}\left(Y_1\right)
=Z^{-1}\exp\left\{ n_0 \D{V} \left[\frac{1}{2}\,\ln\left(1-Y_1\right)\left(Y_1-1\right)
-Y_1\ln\left(Y_1\right)-\frac{1}{2}\, Y_1\left(\ln\left(2\right)-1\right) \right]\right\} ,
\label{eq:P_eq_Hanggi}
\end{equation}
where $Z$ is a normalization constant. This Einstein distribution $P^\mathrm{LM}_{\text{eq}}(Y_1) \sim \exp(S(Y_1)/k_B)$ is in agreement with the
the correct thermodynamic entropy $S$ in the limit of Stirling's approximation, as we demonstrate in Appendix~\ref{AppendixDimerization}.

For the chemical Langevin equation the stochastic production rate for $n_1$ is
\[
\widetilde{\Omega}_1 = -2 \sqrt{\frac{2 \mathcal{D}^\mathrm{CL}}{\D{V}}}\, \mathcal{W}^\Omega
\qquad\mathrm{with}\qquad
\mathcal{D}^\mathrm{CL}
=\frac{1}{2}\left(k^{+}n_1^{2}+k^{-}n_{2}\right).
\]
The equilibrium distribution can also be found from the stationary
solution of the FPE corresponding to the CLE,
which we do not write here for brevity.\footnote{In the literature this equation is called
the CFPE (chemical FPE), see (4) in \cite{CLE_vs_vanKampen}.}
We do note that, for this example, $P^\mathrm{CL}_{\text{eq}}(Y_1)$ is quite close to a Gaussian.
We further observe that, unlike the LME, no thermodynamic interpretation can be given to $\ln\, P_{\text{eq}}^\mathrm{CL}$.
In fact, the tails of $P_{\text{eq}}^\mathrm{CL}$ are quite different from those of
$P_{\text{eq}}^\mathrm{LM}$ and, being nearly Gaussian, the former includes unphysical
values of the concentration (i.e., $Y_1$ is not constrained between 0 and 1).

\begin{figure}
\begin{centering}
\includegraphics[width=0.49\textwidth]{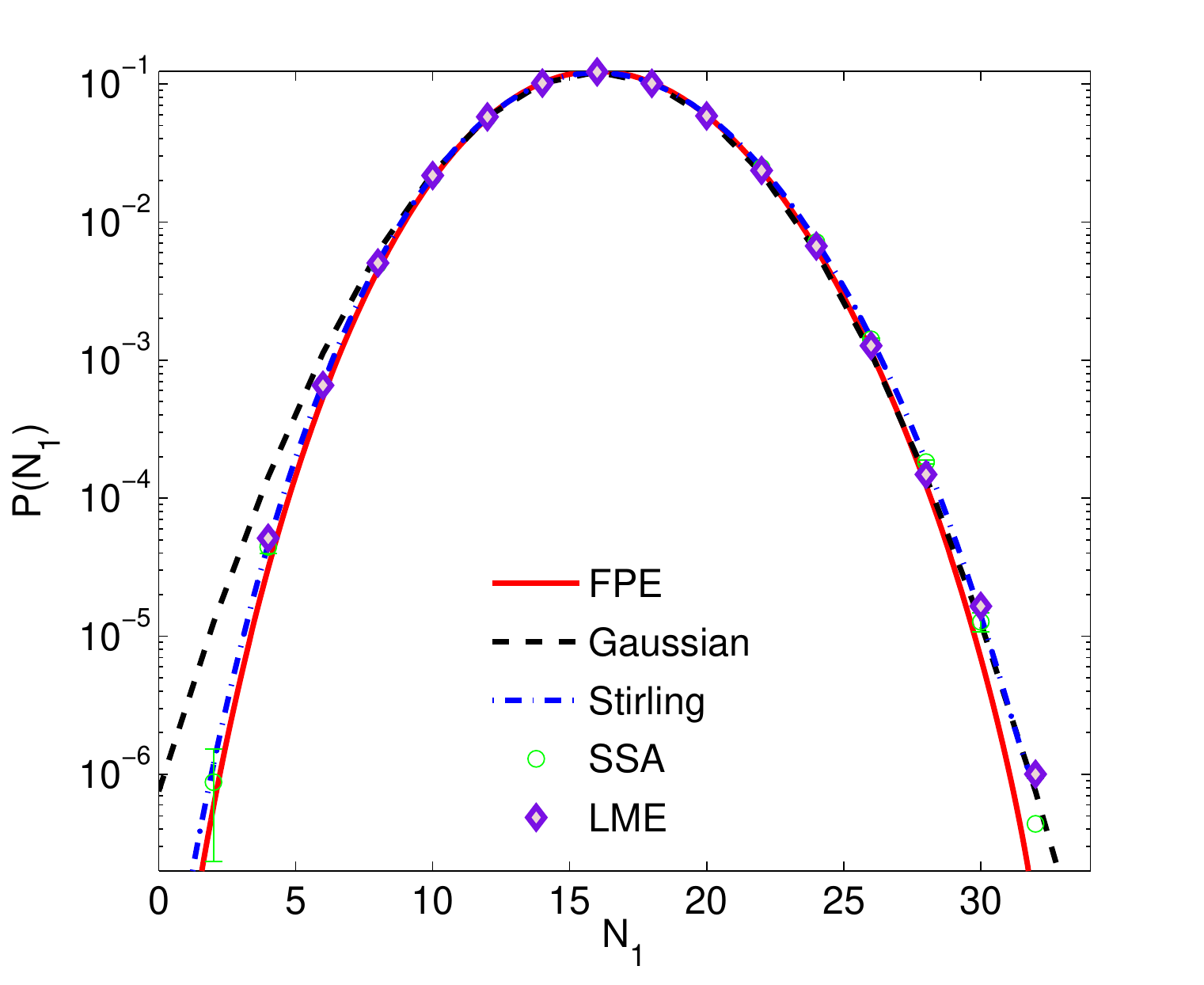}
\includegraphics[width=0.49\textwidth]{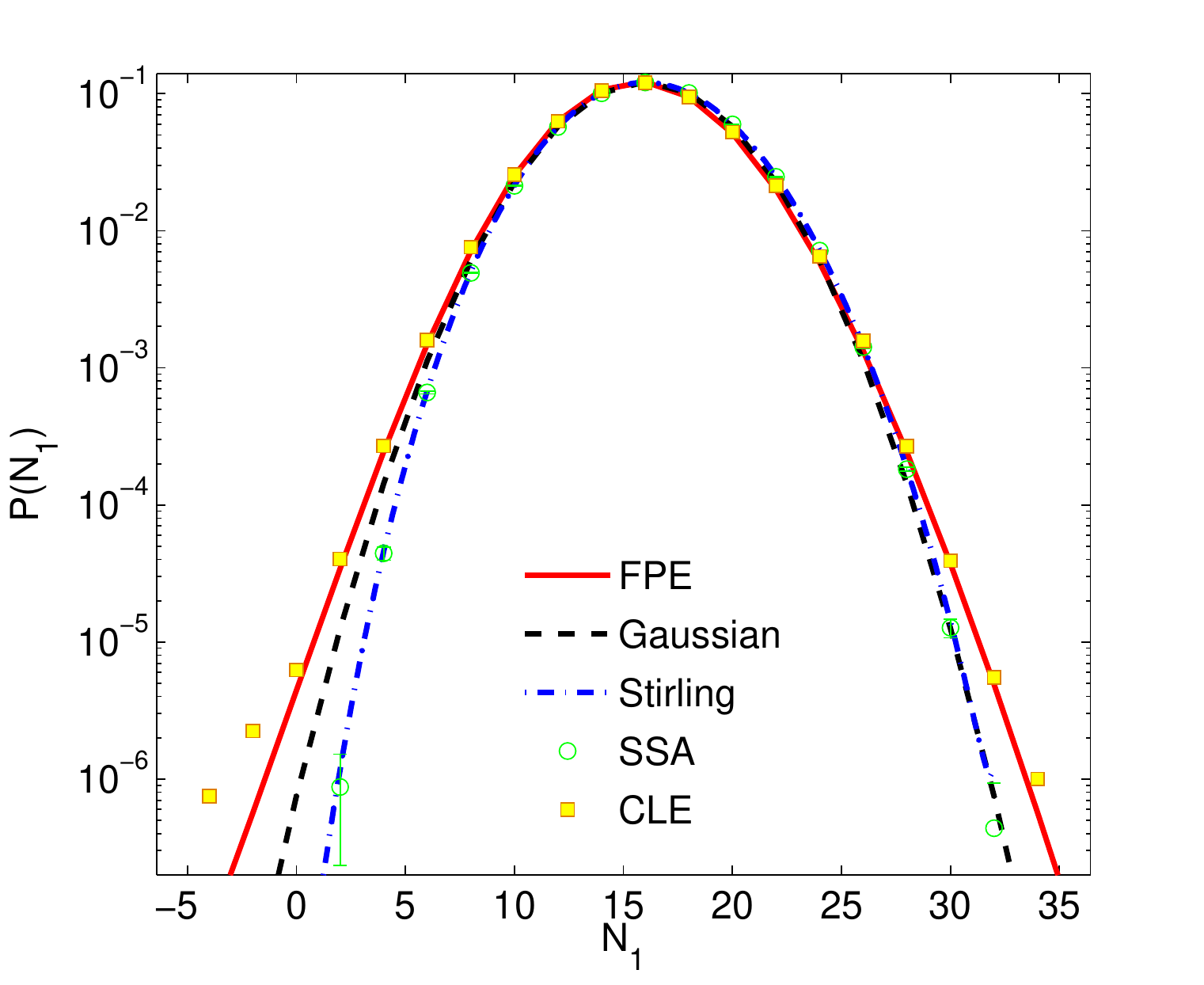}
\par\end{centering}

\centering{}\caption{\label{fig:AA_eq_hist}
Empirical histograms of the probability distribution $P(N_1)$
for the number of monomers $N_1$ obtained from the
log-mean equation (left) and the chemical Langevin equation (right),
compared to results from the CME (obtained using SSA).
The theoretical solution of the corresponding FPE is shown with a solid red line;
the numerical results from SSA are the circles.
The Einstein distributions employing the Gaussian approximation and the second-order Stirling approximation
(\ref{eq:Stirling}) to the true entropy (\ref{eq:counting}) are
also shown. Note that $\langle N_1 \rangle \approx 16$, which is
rather small and thus tests the limits of applicability of a Langevin (non-discrete or non-integer) description.}
\end{figure}

Figure~\ref{fig:AA_eq_hist} shows numerical results for the equilibrium
distribution of the number of monomers $N_1=n_1 \D{V} = \rho Y_1 \D{V}/m_1$.
At thermal equilibrium the simulation results using
the log-mean equation (LME) form for the noise
are in excellent agreement with equilibrium statistical mechanics (see Appendix~\ref{AppendixDimerization})
and with CME/SSA simulation results.
Other work has also shown that, when detailed balance is obeyed, the LME correctly
reproduces the equilibrium transition rates for rare jumps between stable
minima in bistable systems \cite{BistableChemical_FPE}.
On the other hand, the chemical Langevin equation (CLE) result
has the noticeable flaw that, being a Gaussian,
the distribution extends to unphysical negative values of $N_1$.

\begin{figure}
\begin{centering}
\includegraphics[width=0.49\textwidth]{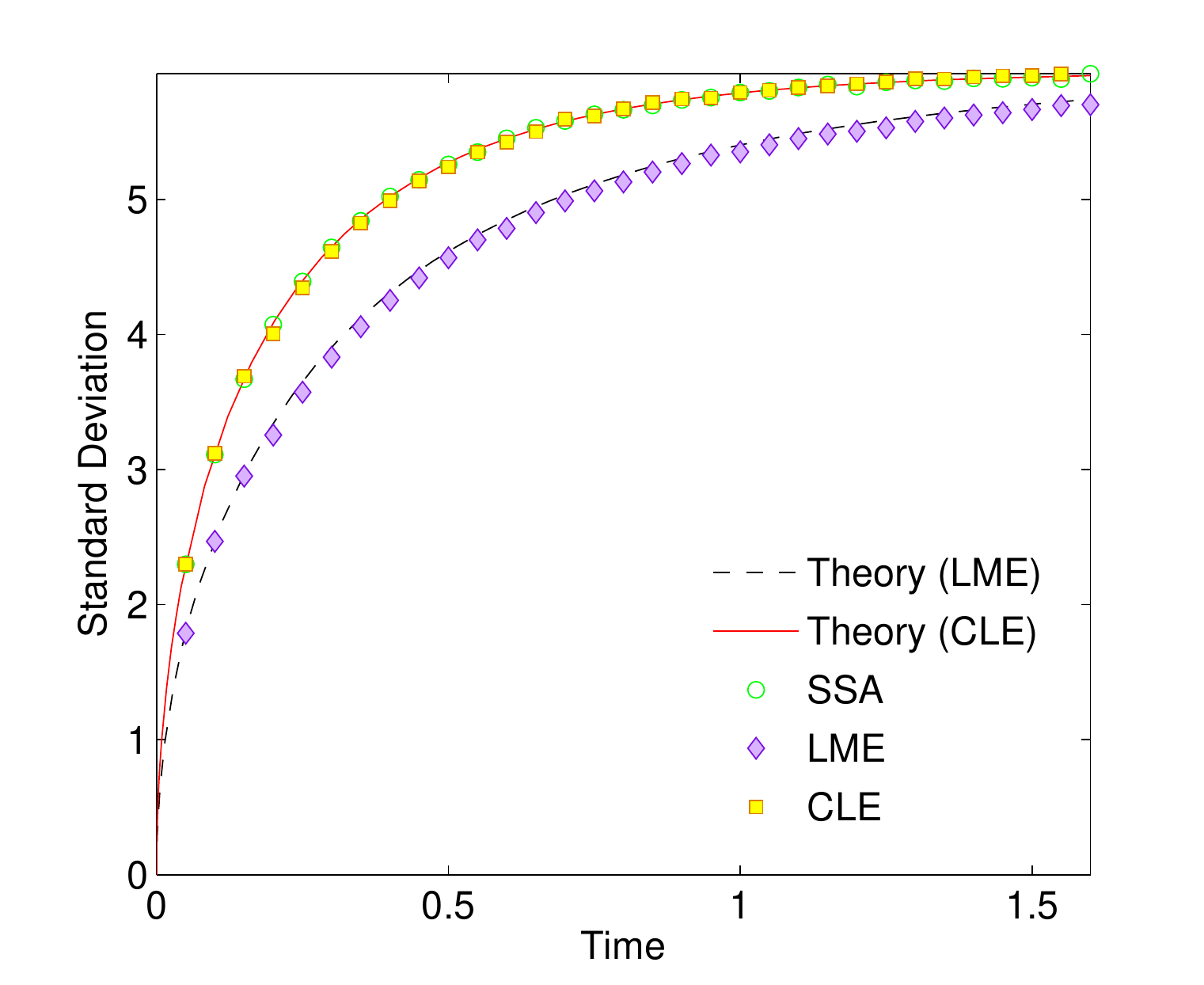}
\includegraphics[width=0.49\textwidth]{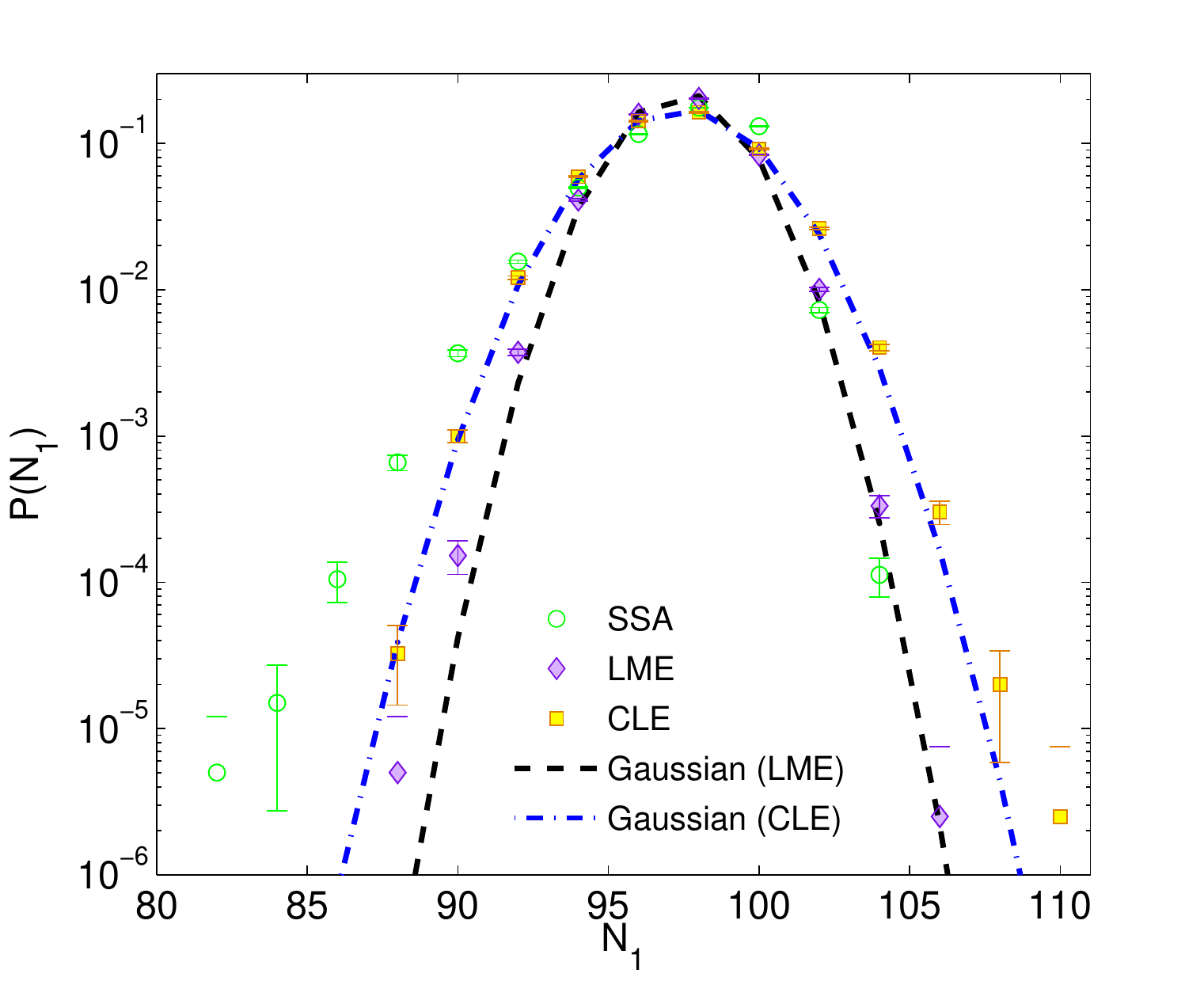}
\par\end{centering}

\centering{}\caption{\label{fig:AA_neq_std}
Relaxation toward equilibrium in a closed
cell initially containing $100$ monomers and $4$ dimers
and relaxing toward the equilibrium state of $\langle N_{1} \rangle \approx 54$ monomers.
Rates are $k^- = 0.3$ and $k^+ = 2.78\cdot 10^{-4}$; time step is $\Delta t = 0.005$.
(Left) Time evolution of the standard deviation of the number of monomers;
theory curves are obtained by solving (\ref{Dimerization_CLT}).
(Right) Histogram of the probability density $P(N_1,\,t)$ for the number of monomers at an early time $t=0.05$. }
\end{figure}

However, the LME does not compare favorably with the numerical solution
of the CME for time-dependent situations,
such as when a system is relaxing toward equilibrium.
To illustrate this, we simulate an ensemble of systems prepared with
an initial condition far from equilibrium,
specifically with $Y_1(t=0) \approx 1$, and measure the time-dependent probability distribution $P(Y_1,t)$
as the system relaxes toward chemical equilibrium ($(Y_1)_\mathrm{eq} = \frac12$).
As expected, for the ensemble mean value of the number density
$\bar{n}_1(t)=\langle n_1(t) \rangle$, we find close agreement
among LME, CLE, and CME results (not shown), even when fluctuations are quite large.
However, if we consider the standard deviation of the number of monomers,
the left panel in Fig.~\ref{fig:AA_neq_std} clearly demonstrates that
the CLE is in much better agreement with the CME (as shown by the SSA results)
for describing relaxation toward equilibrium.
Also shown on this graph is the theoretical solution for the standard deviation
obtained by first linearizing the CLE
\footnote{The linearized CLE is known as the second-order van Kampen expansion \cite{vanKampen:07} in the physics
literature and has the same formal order of accuracy
as the CLE \cite{KurtzTheorem_CLE} but is much simpler to solve analytically due to its linearity.
Analysis in Ref. \cite{CLE_vs_vanKampen} suggests that the nonlinear CLE may be more accurate than
the linearized CLE for large noise but we are not aware of rigorous mathematical estimates.}
around the solution of the deterministic law of mass action
(which is the law of large numbers corresponding to the CME~\cite{KurtzTheorem_CLE}),
and then writing a system of ODEs for the mean and variance of $n_1(t)$.
Specifically, we have that
$d\bar{n}_1(t) / dt = \Omega_1$ and, using Ito's formula,
we get the central limit theorem corresponding to the CME~\cite{KurtzTheorem_CLE},
\[
\frac{dC_1(t)}{dt} = -2 C_1(t) \frac{d\Omega_1 \left(\bar{n}_1(t)\right)}{dn_1} + \frac{8}{\D{V}} \mathcal{D}^\mathrm{CL} \left(\bar{n}_1(t)\right),
\label{Dimerization_CLT}
\]
where $C_1(t) = \langle \left(n_1(t)-\bar{n}_1(t)\right)^2 \rangle$.

The agreement between CLE and CME in the left panel of Fig. \ref{fig:AA_neq_std} is not surprising since
it is well-known that the central limit theorem for the CME
is a linear Langevin equation with the noise covariance given by the
CLE form rather than the LME form.~\cite{KurtzBook,KurtzTheorem_CLE}
The agreement between the CLE and SSA results is less impressive
when we look more closely at the probability distribution
during the relaxation toward equilibrium.
The right panel of Fig.~\ref{fig:AA_neq_std} shows
histograms of the probability distribution for the monomers, $P(N_1,t)$, at an early time in the relaxation.
This distribution is \emph{not} close to Gaussian for the CME/SSA solution
and we see that, in this regard, the CLE result is no better than the LME result.
In fact, for the probability distribution of the dimers (not shown)
the CLE results have the unphysical feature
of non-zero probability for negative values of dimer concentration.
\modified{Of course, one can argue that the number of molecules in the system is too small for a Langevin approximation to apply.
If the fluctuations are decreased, the probability distribution $P(N_1,t)$ will become
closer to Gaussian and then the CLE will provide a better description; note however
that the tails of the distribution will always be incorrect for the CLE, even at thermodynamic equilibrium.}

\subsubsection{Bistable BPM Model}\label{WellMixedBpmSection}

As a less trivial homogeneous example,
we consider the BPM model (\ref{BPMmodelEqn}) for an open system held at a non-equilibrium steady state
for which the probability distribution function is bimodal.~\cite{Baras1996}
For the chemistry-only study in this section,
the number of molecules of species 1, 2, and 3 ($U$, $V$, and $W$)
are allowed to vary while the number of molecules of all other species are fixed.
The relevant parameters are given in Table~\ref{table:BPM_Bimodal}.
Note that a similar system (with all reactions being bimolecular) was studied by Baras et al.,
who found good agreement between SSA and molecular simulations using the
Direct Simulation Monte Carlo (DSMC) algorithm ~\cite{Baras1996}.
Baras et al. also examined the accuracy of the CLE {\em linearized} around the
solution of the deterministic equations, and, not surprisingly, found it to be a very poor approximation
of the CME for the parameters they chose. Gillespie \cite{ChemicalLangevin_Gillespie} suggests
that ``A repetition of the study of Baras and co-workers using
the Langevin equation [CLE] instead of the [linearized CLE]
should show the chemical Langevin equation in a fairer light.''
This section presents such a study using both the CLE and the LME.
The parameters were selected such that the number of particles is large
(roughly $O(10^2-10^3)$ for each species) but not so plentiful as to prevent
the SSA simulation from accurately sampling the bimodal distribution in a reasonable
amount of computation time.

\begin{table}

\begin{tabular}{|l|c|c|c|}
  \hline
  Species & $S$ & $U_f$ & $V_f$ \\
  \hline
  $n_r$ (fixed) & $7.0\cdot10^2$ & $5\cdot 10^{11}$ & $5.65\cdot 10^{10}$ \\
  \hline
\end{tabular}
\smallskip
\begin{tabular}{|l|c|c|c|c|c|}
  \hline
  - & $\mathfrak{R}_1$ & $\mathfrak{R}_2$ & $\mathfrak{R}_3$
  & $\mathfrak{R}_4$ & $\mathfrak{R}_5$ \\
  \hline
  $k^+$ & $2\cdot 10^{-3}$ & $10^{-3}$ & 0.0200936 & 0.28 & 0.28 \\
  \hline
  $k^-$ & $2\cdot 10^{-9}$ & $2.198\cdot 10^{-4}$ & $2.009\cdot 10^{-8}$
  & $2.8\cdot 10^{-10}$ & $2.8\cdot 10^{-10}$ \\
  \hline
\end{tabular}
\caption{\label{table:BPM_Bimodal}
Table of parameters for the BPM model in a well-mixed system (see Figure~\ref{fig:BPM_trajectory}).
The system volume $\D V=4$ and the time step $\D t=1.0$ (reducing
the time step size further did not change the results significantly).
}.
\end{table}

\begin{figure}
\begin{centering}
\includegraphics[width=0.49\textwidth]{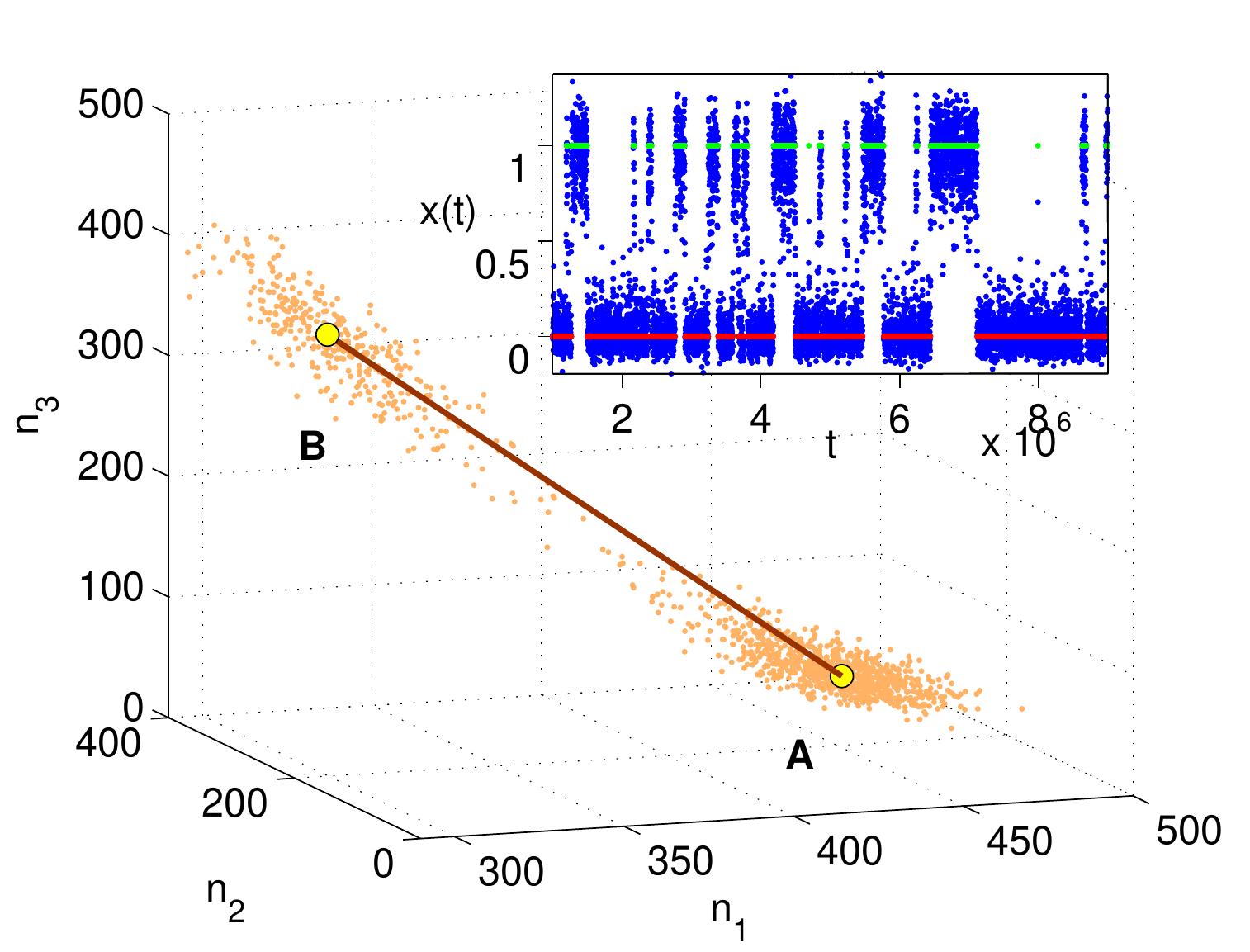}
\includegraphics[width=0.49\textwidth]{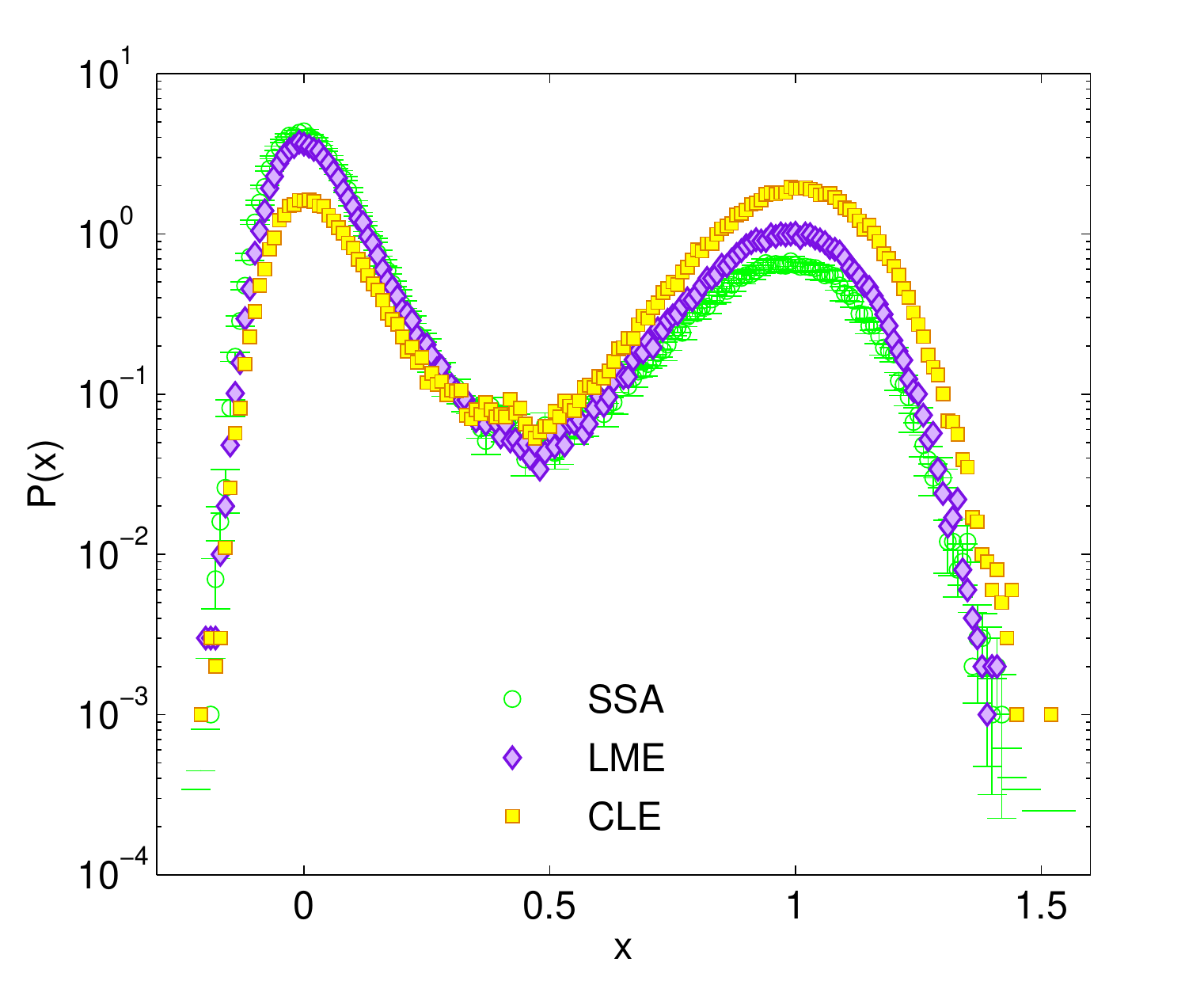}
\par\end{centering}
\centering{}\caption{\label{fig:BPM_trajectory}
Numerical results for the BPM modelin a well-mixed system (see Table~\ref{table:BPM_Bimodal}).
(Left) A phase space picture of a typical trajectory from SSA as
it visits the two basins around the deterministic steady states
(labeled A and B in the figure). The insert shows the trajectory in the collective coordinate, $x(t)$,
based on assigning each point to either state A (red) or state B (green) based on the last state visited by the trajectory (see text).
(Right) Histogram of the steady-state probability distribution $P(x)$ comparing SSA, LME, and CLE simulation results.
Measurements were skipped for the first
$5\cdot 10^5$ steps to relax the initial transient
and then statistics were collected for $1\cdot10^8$ steps.
}
\end{figure}

A phase-space picture of a typical trajectory is shown
in the left panel of Fig.~\ref{fig:BPM_trajectory}.
The trajectory moves between two basins centered around the two stable
deterministic steady states, which are labeled state
$A$ (corresponding to $N_1\approx1740,\, N_2\approx448,\, N_3\approx328$ molecules)
and state $B$ (corresponds to $N_1\approx1224,\, N_2\approx936,\, N_3\approx1424$ molecules).
Based on this picture, we chose to define a collective coordinate
$x(t)$ which is the projection of the state $(n_1, n_2, n_3)$ onto
the line connecting the two stable points (red line in the figure).
This simple linear collective coordinate has the property that $x=0$
at state $A$ and $x=1$ at state $B$; note that $x(t)$ is \emph{not} bounded
between zero and one.
The insert in Fig.~\ref{fig:BPM_trajectory} (left panel) shows $x(t)$
for a typical trajectory as the system moves between the basins.

The right panel in Fig.~\ref{fig:BPM_trajectory} shows the steady-state probability distribution $P(x)$
for the collective coordinate $x$, clearly illustrating its bimodal form;
similar results are found for the probability distributions
for $n_1$, $n_2$, and $n_3$.
The results for the LME and CLE Langevin approximations
are qualitatively similar to those from the CME/SSA but quantitatively different;
the LME result is in better agreement with the CME for this specific example.
To also examine the long-time dynamics of the well-mixed bistable BPM system,
we assign each (discrete) point in time to either state $A$ or state $B$
(see insert in left panel of Fig.~\ref{fig:BPM_trajectory}).
The assignment is performed by defining two sets $\mathcal{A}=\left\{ x<0.3\right\} $
and $\mathcal{B}=\left\{ x>0.6\right\} $ and assigning each point
in the trajectory to the \emph{last} set that was visited. The distribution of
waiting times spent in the two states before transitioning to another state is
related to the transition rate, and the ratio of the average waiting times gives the ratio of the probabilities
to be found in each of the two states. For large $\D{V}$ (weak noise), the
transitions are rare events and the distribution of waiting times should be approximately
exponential (recall that for an exponential distribution the variance is the square of the mean).
Numerical results for the mean and variance of the time spent in state B before transiting to state A, and vice versa, are shown in Table~\ref{AtoB_Table}.

\begin{table}
\begin{tabular}{|l|c|c||l|c|c|}
  \hline
  State A & Mean & Variance & State B & Mean & Variance \\
  \hline
  SSA & 1.13$\cdot 10^5$ & 1.53$\cdot 10^{10}$ & SSA & 2.86$\cdot 10^5$ & 8.74$\cdot 10^{10}$ \\
  LME & 3.23$\cdot 10^5$ & 9.03$\cdot 10^{10}$ & LME & 8.45$\cdot 10^5$ & 1.91$\cdot 10^{11}$ \\
  CLE & 1.38$\cdot 10^5$ & 1.70$\cdot 10^{10}$ & CLE & 5.01$\cdot 10^5$ & 7.91$\cdot 10^{10}$ \\
  \hline
\end{tabular}
\caption{\label{AtoB_Table} Numerical results for the mean and variance of the time spent in state B before transiting to state A, and vice versa, for the three implementations of stochastic chemistry.}
\end{table}

Our results indicate that for the BPM model the CLE and LME provide a
reasonably good qualitative description of the long-time dynamics and rare-event statistics for the parameters studied here.
However, both approximations are in general uncontrolled and
the quantitative match between the CME and either CLE or LME will not improve even if the cell volume increases and the
fluctuations decrease in amplitude. In Ref. \cite{BistableChemical_FPE} it is observed that the LME correctly reproduces the very long-time dynamics (more precisely, the large deviation theory) of the CME for the bistable Schlogl model. This conclusion is, however, specific to this simple one-dimensional model because the system obeys detailed balance even though it is not in thermodynamic equilibrium; the BPM model studied here is not in detailed balance and there is no {\em a priori} reason to expect the LME to be more accurate than the CLE. The fact that the CLE is not able to describe rare events is well-known, see for example the discussion by Gillespie in \cite{CLE_Isomerization} and after Eq. (9b) (which is the CLE) in \cite{ChemistryReview_Gillespie}, or recent numerical studies of noise-induced multistability \cite{Multistability_CLE}.
It can, in fact, easily be proven that this problem is shared by {\em all} diffusion process (SODE) approximations of the CME
\footnote{While Langevin approximations cannot approximate {\em atypical} statistics of the CME, the CLE is likely appropriate for describing the {\em typical} behavior of the CME \cite{CLE_vs_vanKampen}.},
and fundamentally stems from the difference between the rare-event statistics of Gaussian and Poisson noise
\footnote{To see this contrast the quadratic Hamiltonian in (1.4) in \cite{GMA_CME}, which applies to SODEs, and the exponential Hamiltonian in (1.6) in \cite{GMA_CME}, which applies to master equations.}. A promising alternative is to use tau-leaping to approximately integrate the CME in time~\cite{ChemistryReview_Gillespie} since it uses Poisson noise, and thus has the potential to correctly approximate the long-time behavior of the CME. This point is discussed further in the Conclusions.

\subsection{Giant Fluctuations}\label{GiantFluctuationsSection}

We now consider a system for which concentration fluctuations are affected by both
chemistry and hydrodynamics in an interesting fashion.
In the absence of chemistry a gradient of concentration induces
a long-ranged correlation of
concentration fluctuations~\cite{FluctHydroNonEq_Book,DiffusionRenormalization_PRL,DiffusionRenormalization}.
These correlations are closely tied to the experimentally
observed ``giant fluctuation'' phenomenon \cite{GiantFluctConcentration_Sengers,GiantFluctuations_Nature,FractalDiffusion_Microgravity}.
In an isothermal, nonreacting binary mixture the static structure factor
for fluctuations in the mass fraction of the first species contains two contributions,
\[
S\left(k\right)=\av{(\widehat{\delta Y_1})(\widehat{\d Y_1})^{\star}}=S_{\text{eq}}+S_{\text{neq}},
\]
where ``hat'' denotes a Fourier component;  the equilibrium part is
\begin{equation}
S_{\text{eq}}=\frac{m_1}{\rho}(Y_1)_{\text{eq}}\left(1-(Y_1)_{\text{eq}}^{2}\right),\label{eq:S_AA_eq}
\end{equation}
The non-equilibrium enhancement of
the static structure factor due to a concentration gradient is
$S_\mathrm{neq}\left(k\right) \sim \left(\nabla Y_1\right)^{2} / k^{4}$,
where the wavevector $\V{k}$ is perpendicular to the imposed concentration gradient.

The nature of these long-ranged correlations is modified in the presence of
chemical reactions, as predicted by linearized
fluctuating hydrodynamics ~\cite{LLNS_ReactionDiffusion_S_k,LLNS_ReactionDiffusion,LLNS_ReactionDiffusion2}.
Some preliminary numerical studies of fluctuations in the presence of chemistry
have been done in Ref. \cite{GiantIsomerization_Zarate} using an RDME-based description. However, these simulations
are for a simpler isomerization $ A\rightleftarrows B$ in one dimension and, furthermore,
they are concerned with reaction-diffusion only and do not
account for the hydrodynamic velocity fluctuations
that are responsible for the giant concentration fluctuation phenomenon.

We consider here the dimerization reaction (\ref{DimerizationEqn}) in a spatially inhomogeneous system.
A rather detailed linearized fluctuating hydrodynamic theory for this example
has been developed by Bedeaux et al. in \cite{LLNS_ReactionDiffusion2},
for a system in which a concentration gradient is imposed via a temperature gradient through the Soret effect.
However, this analysis assumes a liquid mixture (large Schmidt and Lewis numbers) and thus does not apply to gas mixtures.
Therefore, a simplified theoretical analysis of giant fluctuations in binary gas mixtures in the presence of
an imposed constant concentration gradient and reactions is developed in Appendix~\ref{AppendixGiant}.

Our simplified theory decouples the temperature equation and uses a concentration
equation (specifically the mass fraction of the first species) coupled
to an incompressible fluctuating velocity equation.
For the case of a {\em liquid} mixture with very large Schmidt number,
which is the case considered in \cite{LLNS_ReactionDiffusion2}, the calculation predicts that the nonequilibrium enhancement of the
static structure factor of concentration fluctuations for the monomer species is
(see eqn.~(\ref{eq:S_AA_simple}))
\begin{equation}
S_\mathrm{neq}\left(k\right)=
\frac{k_{B}T\left(\nabla Y_1\right)^{2}}{\eta D k^{4}}\,\left(1+\left(d k\right)^{-2}\right)^{-1},
\label{eq:S_AA}
\end{equation}
where $D$ is the diffusion coefficient and $\eta$ is the viscosity.
The last term on the r.h.s. depends on the penetration depth $d$~\cite{DM_63},
\[
d=\sqrt{\frac{D}{3k^{-}}}.
\]
We see that for large wavenumbers ($k\gg 1/d$) the spectrum is $\sim k^{-4}$,
as in the absence of the chemical reaction.
However for small wavenumbers ($k \ll 1/d$)
there is a transition to a $k^{-2}$ spectrum. For gas mixtures, however,
a more detailed model is required that takes into account the finite value of the Schmidt number.
The result of this calculation is eqn.~(\ref{eq:S_AA_rnertial}), which predicts a further transition to a
flat (constant) spectrum at small wavenumbers, with a finite
$S_{\text{neq}}(k=0)=(k_{B}T) \left(\nabla Y_1\right)^{2}/(9 \rho (k^-)^2)$.
The calculation in Appendix~\ref{AppendixGiant} indicates that this effect is important even in liquids
and the more refined theory ought to be used if quantitative agreement with experiments or simulations is sought.

We performed a series of simulations to investigate these predictions
using the full hydrodynamic equations plus chemistry.
It is important to note that even though we use the full nonlinear equations,
nonlinearities in the fluctuations are negligible for the simulations reported here
\footnote{The nonlinearity of the deterministic (macroscopic) equations is fully taken into account}.
In fact, the noise is very weak (since the domain is quite thick in the $z$ direction) and
the numerical method is effectively performing a computational linearization
of the fluctuating equations around the solution of the (nonlinear) deterministic equations \cite{MultiscaleIntegrators};
this is simular to what is done analytically in Refs. \cite{LLNS_ReactionDiffusion,LLNS_ReactionDiffusion2} but does
not require any approximations.
In the small Gaussian noise regime the linearized CLE equation applies, and therefore in these simulations we use the CLE
form for the stochastic chemical production rate in agreement with the theory in \cite{LLNS_ReactionDiffusion,LLNS_ReactionDiffusion2}.
Identical results (to within statistical errors) are obtained by using the LME form of the noise (not shown);
this is not unexpected since
the important noise here is the stochastic momentum tensor driving the velocity fluctuations;
the stochastic mass flux and production rates only affect the reaction-diffusion part of
the spectrum, which is much smaller than the nonequilibrium enhancement we study here.

Here we assume that the traditional number-density based LMA (\ref{eq:LMA_numdens}) holds
with constant rates $k^+$ and $k^-$. This requires that the time scale for
the reaction is proportional to the number density, that is, $\tau \sim p/k_B T = n$.
From (\ref{eq:eq_const}) and (\ref{eqn:mu_rdealgas}),
for the dimerization of an ideal gas the ratio of these rates is,
\[
\frac{k^{+}}{k^{-}} = \frac{(\Lambda_1^3)^2}{\Lambda_2^3}\frac{j_2}{(j_1)^2}
= 2^{3/2} \Lambda_1^3 \frac{j_2}{(j_1)^2}
\]
which is a complicated function that depends on the form of
the internal degrees of excitation.
These details determine the number fraction $(X_{1})_\mathrm{eq}$
or, equivalently, the mass fraction $(Y_{1})_\mathrm{eq}$ at chemical equilibrium;
here we set the ratio of the forward and reverse rates to ensure
$(Y_{1})_\mathrm{eq}=1/2$, assuming that $\Lambda_1$, $j_1$ and $j_2$ are consistent with this choice.
Since the reaction here changes the number density and thus the pressure,
the reaction is strongly coupled to the momentum and energy transport equations.
In order to minimize this coupling, we adjust
the number of internal degrees of freedom of the dimer particles. Specifically, we set
the heat capacities to $c_{p,1} = \frac52 k_B/m_1$ (corresponding to three translational
and zero internal degrees of freedom),
$c_{p,2} = 5 k_B/m_2$ (corresponding to three translational and five internal degrees of freedom). This choice
ensures that $c_p$ of the mixture is independent of composition so that at constant pressure the
reaction is isothermal.

The fluid was taken to be a dilute binary mixture of hard-sphere gases,
using kinetic theory formulae for the transport coefficients \cite{Bell_10}.
In CGS units, the species diameters are $\sigma_{1}=2.58\cdot10^{-8}$ and $\sigma_{2}=3.23\cdot10^{-8}$,
and $m_{1}= 6.64 \cdot 10^{-23}$.
At equilibrium the density $\rho_\mathrm{eq}=1.78\cdot10^{-3}$,
the temperature $T_\mathrm{eq}=300$,
and the concentration $(Y_{1})_\mathrm{eq}=0.5$.
For these parameters, at the equilibrium conditions the mass diffusion
coefficient is $D=0.2698$
while the momentum diffusion coefficient (kinematic viscosity) is $\nu=0.2374$.
The ratio of the reverse and forward reaction rates is fixed at $k^{-}/k^{+}=\rho/m_1=2.67985\cdot10^{19}$.
We vary the penetration depth $d$ by changing the value of the reaction rates.

\begin{figure}[ht]
\begin{centering}
\includegraphics[width=0.75\textwidth]{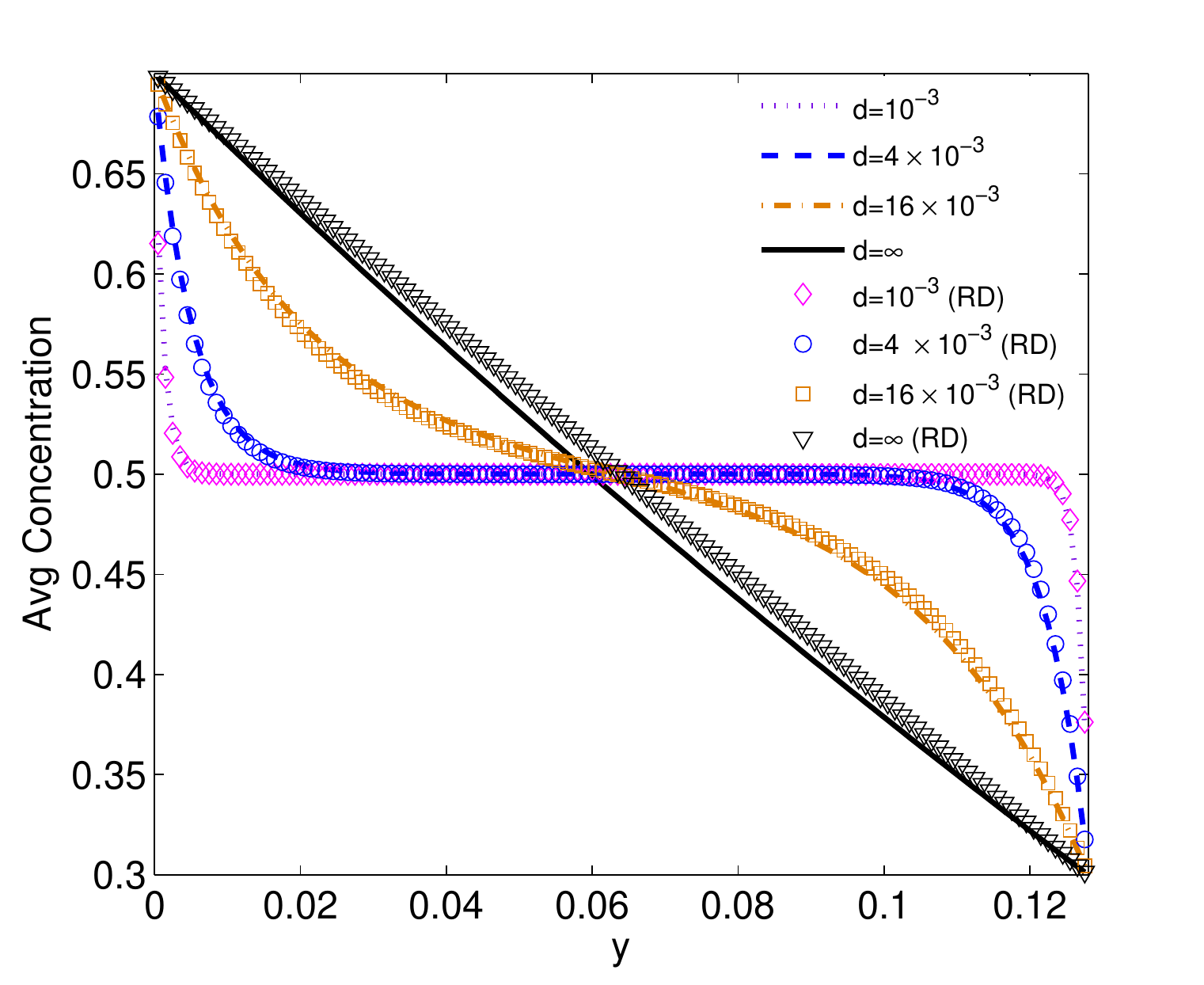}
\par\end{centering}

\centering{}\caption{\label{fig:AvgConc}
Average concentration profile $Y_1(y)$ at steady
state for various values of penetration depth, $d$.
Lines are from simulations using the full hydrodynamic equations;
symbols are for reaction-diffusion only.}
\end{figure}

The simulations used a $128^{2}$ grid and
time step size $\D t=2.5\cdot10^{-8}$, grid spacing $\D x=\D y=10^{-3}$,
and thickness in the $z$ direction of $\D z=10^{-3}$.
The first $6\cdot10^{4}$ time steps were skipped
and then statistics collected for $2\cdot10^{6}$ time steps.
A steady concentration gradient was imposed by using Dirichlet
boundary conditions at top and bottom boundaries ($y=L$ and $y=0$).
Specifically, we take $Y_1\left(y=0,t\right)=0.3$ and $Y_1\left(y=L_{y},t\right)=0.7$,
with temperature fixed at $T=300$
and no-slip boundary conditions for the velocity.
Periodic boundary conditions were used in the other direction.
Concentration profiles for various values of penetration depth, $d$,
are shown in Fig.~\ref{fig:AvgConc}.
As expected, when the chemistry is slow ($d \gg L$)
the concentration profile is nearly linear;
when the chemistry is fast the concentration is nearly constant
(at its chemical equilibrium value of
$(Y_{1})_\mathrm{eq} = 1/2$) except near the boundaries.
Note that we set the thermal diffusion ratio
to zero (i.e., no Soret effect) so that the system is isothermal and the simple
theory presented in Appendix~\ref{AppendixGiant} applies.

For comparison, we also performed simulations
in which we turn off all hydrodynamics except Fickian diffusion, giving us a reference
reaction-diffusion structure factor $S_{rd}(k)$.
For the case of a binary mixture \cite{Bell_10} with $k^{-}/k^{+}=\rho/m_1$, the reaction-diffusion CLE reduces to
\begin{eqnarray}
\frac{\partial }{\partial t} \left( Y_1 \right) &=& \grad\cdot \left( D  \grad Y_1 \right)
  + k^- \left( -2 Y_1^2 + (1-Y_1) \right)     \\
  &+& \grad\cdot\left[ \sqrt{\frac{2 D}{n_0} Y_1 (1-Y_1^2)} \;\ZSpeciesFlux \right]
   -\sqrt{\frac{2 k^-}{n_0} \left( 2 Y_1^2 + (1-Y_1) \right) } \;\ZOmega,
   \label{eq:DimerizationRD}
\end{eqnarray}
where $n_0=\rho_1/m_1 + 2 \rho_2/m_2$ is the total number density of A particles contained in both
monomers and dimers,
which is spatially constant in this reaction-diffusion approximation.
It is important to note that the reaction-diffusion model (\ref{eq:DimerizationRD})
is thermodynamically inconsistent because it ignores the coupling of the chemistry to the energy and momentum transport.
It is well-known that adding chemistry should not change the fluctuations
at thermodynamic equilibrium \cite{LLNS_ReactionDiffusion}, and this is indeed the case for the
complete set of hydrodynamic equations that we study in this work.
By contrast, for the reaction-diffusion (\ref{eq:DimerizationRD}) the equilibrium structure factor
for $(Y_{1})_\mathrm{eq}=1/2$ is given by
\begin{equation}
S_{\text{eq}}^{(\text{rd})}=\frac{3 m_1}{8 \rho} \frac {8/9+ k^2 d^2} {1+ k^2 d^2} ,
\label{eq:S_AA_rd}
\end{equation}
which only approaches the thermodynamically correct answer (\ref{eq:S_AA}) for $kd \gg 1$.
Note that the inconsistency between full hydrodynamics and reaction-diffusion is not evident in Eqs. (27a,28) in \cite{LLNS_ReactionDiffusion},
because the authors of that work ``neglect the dependence of the specific Gibbs energy difference on pressure.''
This inconsistency is not of any importance in our study because we only use the reaction-diffusion
simulations to obtain a baseline to subtract from the full hydrodynamic runs at large wavenumbers; at small wavenumbers
the nonequilibrium enhancement is many orders of magnitude larger than the difference between (\ref{eq:S_AA}) and (\ref{eq:S_AA_rd}).

\begin{figure}[ht]
\begin{centering}
\includegraphics[width=0.65\textwidth]{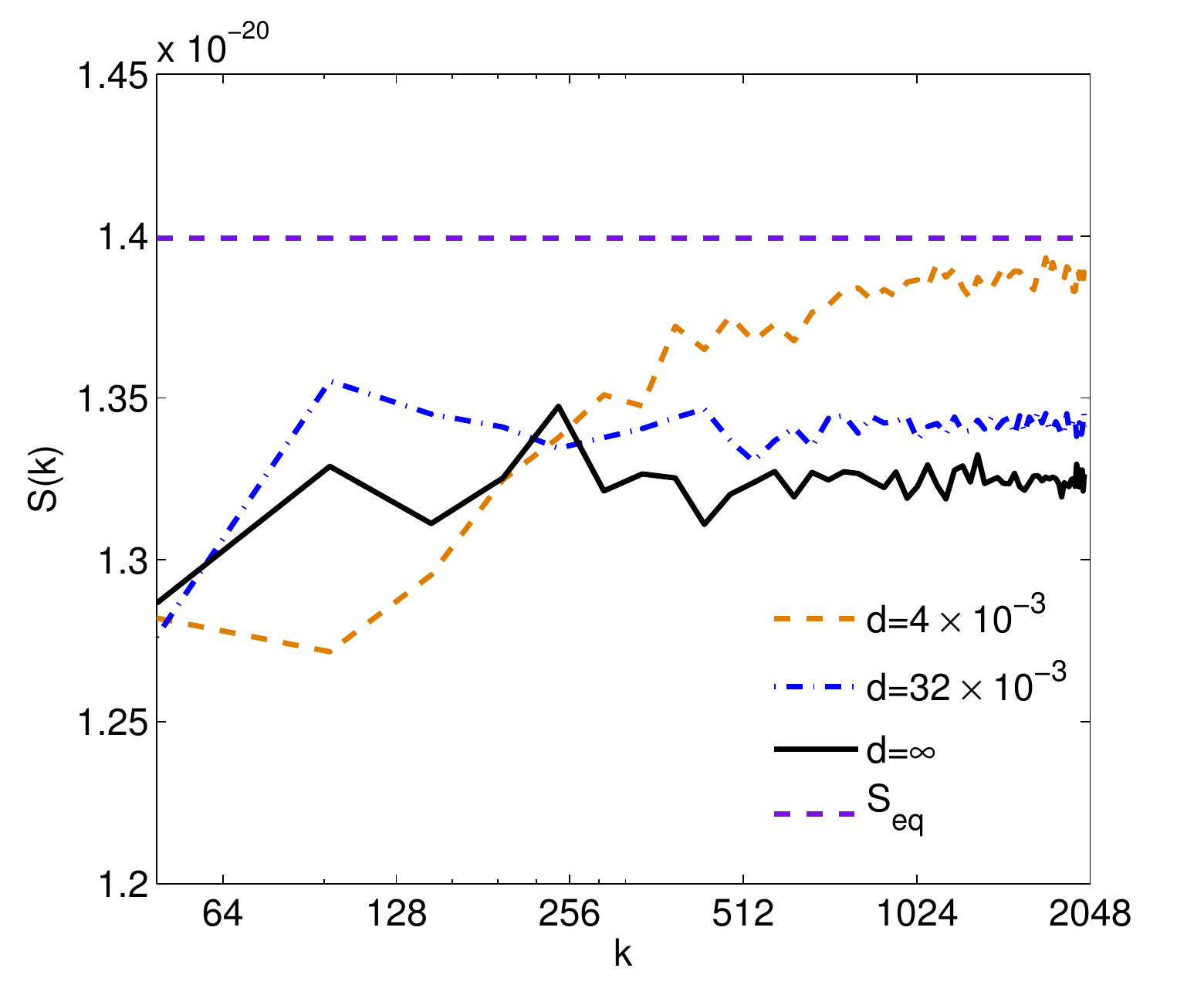}

\includegraphics[width=0.65\textwidth]{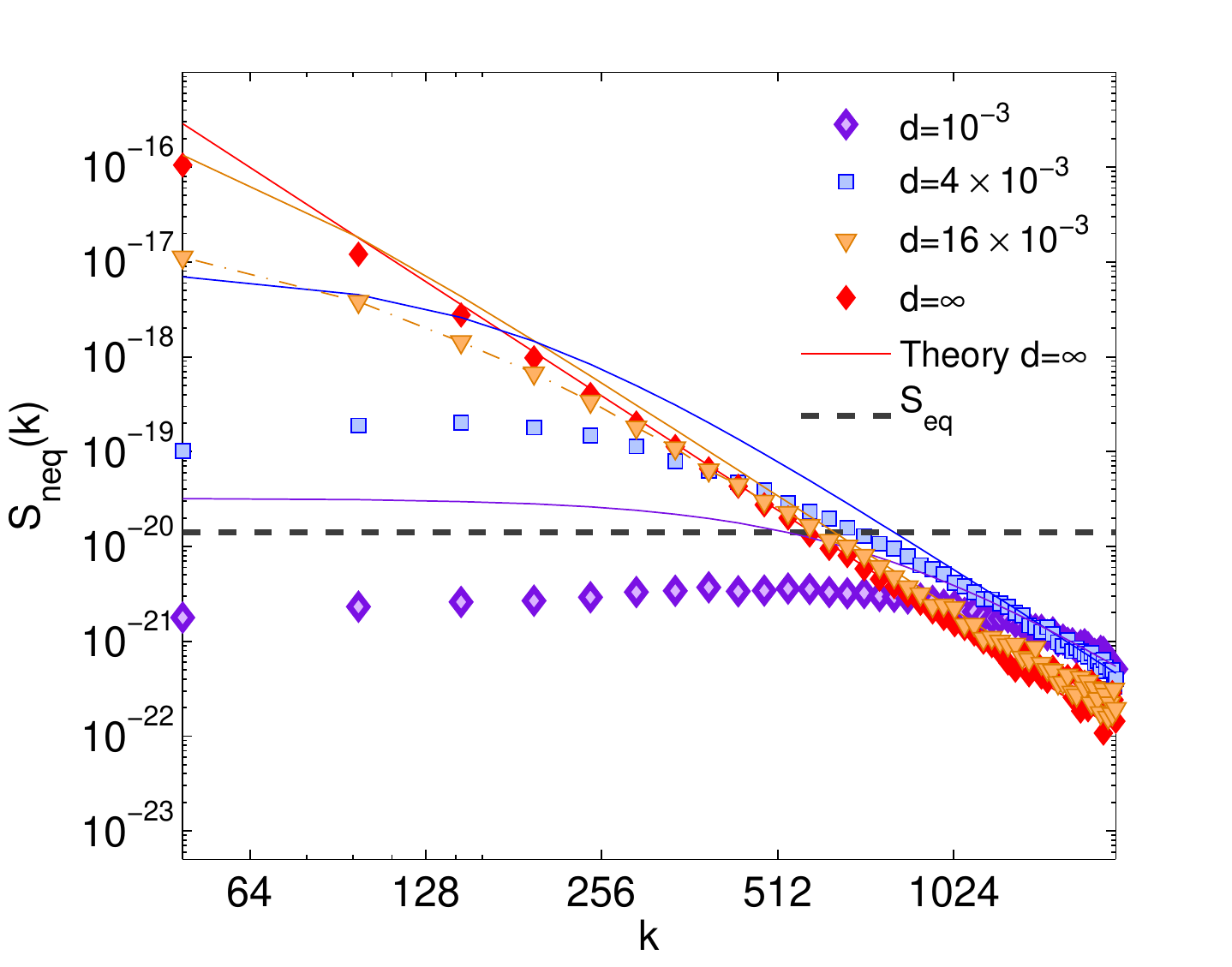}
\par\end{centering}

\centering{}\caption{\label{fig:giant_S_k}
(Top) The structure factor $S_{rd}(k)$ for the reaction diffusion model (\ref{eq:DimerizationRD}) in the presence
of an applied  gradient. For small $d$ (fast reaction) the fact that the structure factor is not perfectly flat (constant)
can be explained by (\ref{eq:S_AA_rd}) and comes from the thermodynamic inconsistency of the reaction-diffusion model.
(Bottom) Non-equilibrium structure factors $S_{\text{neq}}(k)=S(k)-S_{rd}(k)$ for values of
penetration depth $d$ varying from $\infty$ (no reaction) to $d=10^{-3}$, the width of one grid cell.
Numerical results are shown with symbols,
and the theoretical prediction (\ref{eq:S_AA_rnertial}) is shown as a line with the same color.
In the absence of a gradient is flat, $S_{eq} \approx 1.4\cdot10^{-20}$.}
\end{figure}

The reaction-diffusion runs are not limited by the
Courant condition so we increased the time step to $\D t=2.5\cdot10^{-7}$; a total
of $6\cdot10^{4}$ steps were skipped initially and then statistics
were collected for $3\cdot10^{6}$ steps.
As seen in Fig.~\ref{fig:AvgConc} the average concentration profiles
are nearly the same whether the simulations used the full hydrodynamic
equations or simply species diffusion.
For the structure factor, however, we find that
the reaction-diffusion simulations do \emph{not} reproduce the
giant fluctuation result (\ref{eq:S_AA}), rather, they follow (\ref{eq:S_AA_rd}),
which does not show a power-law enhancement, as seen in the top panel of Fig. \ref{fig:giant_S_k}.
This is expected since the giant fluctuation effect arises from the
coupling of concentration fluctuations with the velocity fluctuations.

Because the equilibrium structure factor (\ref{eq:S_AA_eq})
is derived assuming a uniform bulk state, which is not actually the case here,
we define $S_{\text{neq}}(k)=S(k)-S_{rd}(k)$
as a measure of the ``giant'' nonequilibrium
fluctuations coming from the coupling with the velocity equation.
Results for $S_{\text{neq}}(k)$ for several penetration depths are shown in the bottom panel of
Fig. \ref{fig:giant_S_k}, comparing simulation results with the simple theory,
eqn.~(\ref{eq:S_AA_rnertial}).
Since chemistry should have minimal effects for large $k$ according to the theory,
we compute an effective concentration gradient by approximately matching the tail of the numerical
result to the tail of the theory. We see that the theory correctly
reproduces the qualitative trends, namely, that the giant fluctuations
level off to a constant value at a wavenumber of order $d^{-1}$. However, except for the case
of no reaction ($d \rightarrow \infty$),
\footnote{Note that in this case the mismatch between the theory and simulations
at very small $k$ is due to the effect of boundaries \cite{GiantFluctFiniteEffects}.}
the theory is not in quantitative agreement with the simulations.
To confirm that the issue is not under-resolution of
the penetration depth by the grid, we perform runs with a
finer grid of $256^{2}$ cells
\footnote{For the more resolved runs $\D x=\D y=5\cdot10^{-4}$,
$\D t=6.25\cdot10^{-8}$ for the problem without hydrodynamics and
$\D t=6.25\cdot10^{-9}$ with hydrodynamics.}
, and we get the same result over the common
range of wavenumbers, showing these runs are sufficiently resolved for the
purpose of computing $S(k)$. Note that in the plots the numerical wavenumber $k_x$
is corrected to account for discretization artifacts in the standard 5-point Laplacian,
$k^2 = \sin^2(k_x \D{x}/2) / (k_x \D{x}/2)^2$.

The mismatch between theory and simulation is not so surprising since the theory is for a weak gradient
applied to a system that is essentially near equilibrium; this is
not true in this setup. The only way to get this isothermal system out of
equilibrium is via the boundaries, so the system is actually far from
chemical equilibrium near the boundaries and then goes to chemical
equilibrium in the middle of the domain, but in the middle the gradient disappears.
A new more sophisticated theory is required that linearizes not around
a constant state but rather around a spatially-dependent state (this
is automatically done in our code). Also, boundaries (i.e., confinement effects)
may need to be included, especially for penetration depths comparable
to the system size.

\subsection{Pattern Formation}\label{PatternFormationSection}

Since the seminal work of Turing~\cite{Turing1952},
pattern formation in deterministic reaction-diffusion systems has been investigated extensively,
mostly in theoretical studies but also by laboratory experiments~\cite{RD_PatternFormation_Review,cross2009pattern}.
The study of stochastic systems is more recent and, as described in the introduction,
has primarily focused on models based on a
reaction-diffusion master equation (RDME).
Such a model was introduced by Mansour and Houard~\cite{SpatialSSA_Malek} as a
practical numerical scheme for the study of correlations in spatially-distributed reactive systems.
Subsequently RDME-based models have been used to study the influence of fluctuations on
pattern formation for a variety of reaction-diffusion
systems~\cite{Turing_Fluctuations1,Turing_Fluctuations2,Turing_RDME,Turing_RDME_2}.
Recently, a spatial chemical Langevin formulation (SCLE) was proposed~\cite{Spatial_CLE}
and the RDME, SCLE and deterministic equations were compared
for the development of patterns in the Gray-Scott model; it was observed
that the SCLE is qualitatively similar to the RDME for the majority of
examined sets of parameters, but not always.
All of the RDME-based models usually use simplified descriptions of diffusion, but recently it has been observed
that accounting for cross-diffusion effects (which are included in complete generality in our formulation)
may lead to qualitatively-different behavior for Turing instabilities \cite{Turing_CrossDiffusion1,Turing_CrossDiffusion2,Turing_CrossDiffusion3,Turing_CrossDiffusion4}.
Particle simulations including full hydrodynamics have also been performed using
the DSMC method \cite{Turing_DSMC} and molecular dynamics \cite{Turing_MD}; these are,
however, limited to small systems in (quasi) one dimension because of the high computational cost of particle simulations.

In our final example we consider pattern formation in
the Baras-Pearson-Mansour (BPM) model (\ref{BPMmodelEqn})
for a dilute gas mixture with full hydrodynamic transport.
The system is initialized in a uniform constant ``reference'' state in which the number
densities of the different species are as specified in Table~\ref{table:BPM_Pattern}.
These number densities and the reaction rates are set so that
the reference state is similar to that investigated in~\cite{Baras1990}.
Specifically, under the assumption that the number densities of the
reservoir or ``solvent'' species $S$, $U_f$, and $V_f$ are fixed,
the deterministic dynamics of the three reactive species
$U$, $V$, and $W$ starts close to the single unstable fixed point; the stable
attractor of the dynamics is a limit cycle around this unstable point.
Of course, when the number densities of the solvent species are not fixed the dynamics is
six-dimensional and much more complex.
Since we consider a time-dependent non-equilibrium scenario the chemical Langevin form of the noise
(\ref{eq:CLE_Omega_Discrete}) was used for the stochastic chemistry.

In the standard BPM model the solvent species ($S$, $U_f$, and $V_f$)
have fixed concentrations but in our hydrodynamic simulations they were fixed only at the boundaries.
These three species are also made abundant to buffer them from having rapid variations
in concentration (see Table~\ref{table:BPM_Pattern}).
In the limit of infinite concentrations of the solvent species ($S$, $U_f$, and $V_f$) the dynamics
approaches a reaction-diffusion model in which advection as well as momentum and heat transport become negligible.
The reference (initial) values for mole fraction are used to prescribe
Dirichlet boundary conditions for species on each side of the
domain.
This setup mimics open reservoirs in the form of permeable membranes \cite{LowMachExplicit};
note however that implementing boundaries that are also open for advective mass transport
(i.e., inflow and outflow) is quite challenging~\cite{Delgado:08} and not presently supported in our implementation.
At the boundaries, the temperature is fixed at $T = 300$K
and the fluid velocity is set to zero (no-slip)
so species transport is primarily due to mass diffusion.

In our hydrodynamic simulations the fluid is modeled as a hard sphere dilute gas so the transport
coefficients depend on the masses and diameters of the particles of each species.
The particles for all species in the BPM model have equal mass ($m = 6.64 \cdot 10^{-26}$~g)
so as to ensure that the reactions conserve mass.
For all species the particles have only translational energy
and no internal degrees of freedom (i.e., $z=0$) so pressure and enthalpy
are unaffected by reactions.
In the BPM model species $U$ plays the role of the ``inhibitor'' while species $V$ is
the ``activator.''~\cite{cross2009pattern}
Typically pattern formation occurs when the inhibitor diffuses faster than the activator
so we set the diameter of species $U$ particles to be smaller than that of $V$ particles,
specifically $d_1 = 0.125$~nm and $d_2 = 0.5$~nm.
Since we take $k_1^+ \gg k_1^-$ (see Table~\ref{table:BPM_Pattern}),
the intermediary species, $W$, supports the activation of $V$ and
thus we set the diameters of $V$ and $W$ to be equal
(this makes these two species hydrodynamically indistinguishable).
The diameters of the other species ($S$, $U_f$, and $V_f$)
are small, $d_4 = d_5 = d_6 = 0.025$~nm, so they diffuse rapidly from the boundaries and within the system.
These specific values of the diameters are chosen such that
the self-diffusion coefficient of $S$ (and $U_f$, $V_f$) is roughly an order of magnitude
larger than the diffusion coefficient of $U$, which is itself an
order of magnitude larger than the diffusion coefficient of $V$ (and $W$).

The system is simulated in a
rectangular domain that is divided into $128 \times 128 \times 1$ cells
with $\Delta x = \Delta y = 100$~nm.
The magnitude of the noise is varied by varying the domain thickness,
which was either $\Delta z = 100$~nm (low noise) or $10$~nm (high noise).
The reference value for the number of molecules per cell for the
species of interest, $U$, is $O(10^4)$ in the former case and $O(10^3)$ in the latter.
The total number of solvent molecules per cell is $O(10^6)$
for weak noise and $O(10^5)$ for strong noise.
The time step is $\Delta t = 10$~ps, as determined from stability requirements for the explicit temporal integrator.

\begin{figure}
\begin{centering}

\includegraphics[width=0.19\textwidth]{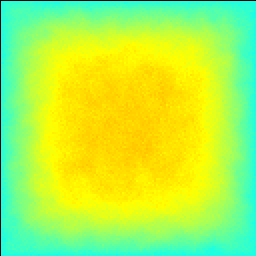}
\includegraphics[width=0.19\textwidth]{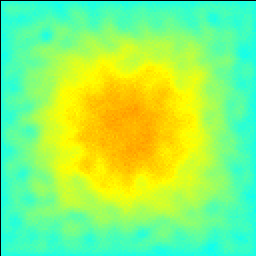}
\includegraphics[width=0.19\textwidth]{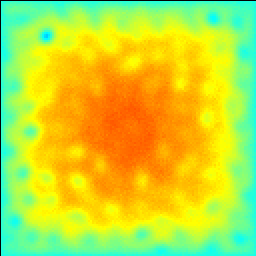}
\includegraphics[width=0.19\textwidth]{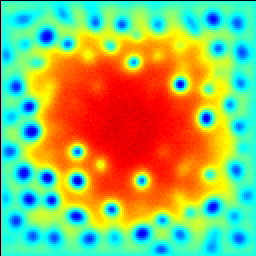}
\includegraphics[width=0.19\textwidth]{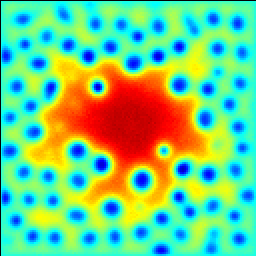}

\includegraphics[width=0.19\textwidth]{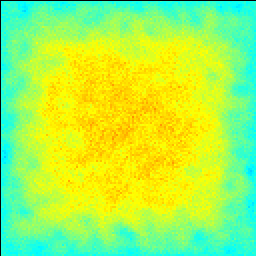}
\includegraphics[width=0.19\textwidth]{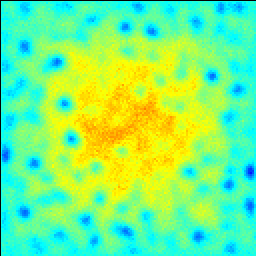}
\includegraphics[width=0.19\textwidth]{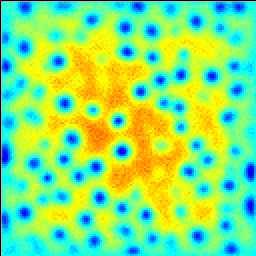}
\includegraphics[width=0.19\textwidth]{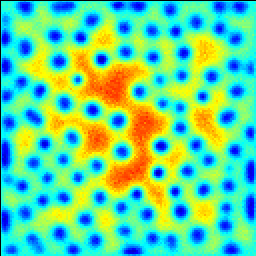}
\includegraphics[width=0.19\textwidth]{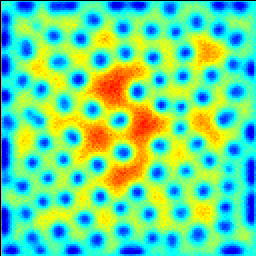}

\includegraphics[width=0.19\textwidth]{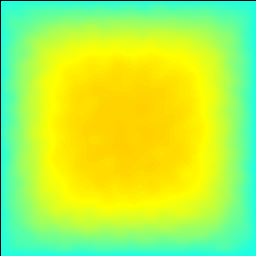}
\includegraphics[width=0.19\textwidth]{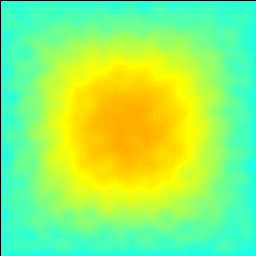}
\includegraphics[width=0.19\textwidth]{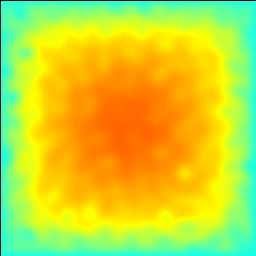}
\includegraphics[width=0.19\textwidth]{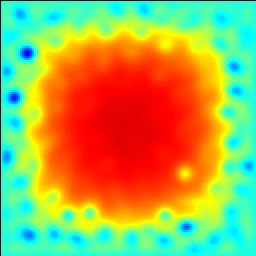}
\includegraphics[width=0.19\textwidth]{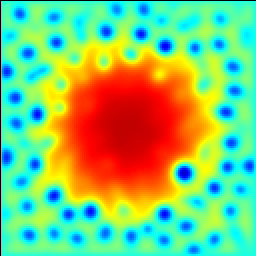}

\par\end{centering}

\centering{}\caption{\label{fig:turing_hom}
Time sequence of images of $\rho_1$ (species $U$) at
$t=10$, 15, 20, 25, and 30 $\mu$s
(time step $\Delta t = 10$~ps,
for other parameters see Table~\ref{table:BPM_Pattern}).
The top row is
a low noise case corresponding to a domain of thickness of $100$~nm;
the middle row is
a high noise case corresponding to a domain of thickness $10$~nm;
and the bottom row is the deterministic evolution with an initial perturbation
introduced by keeping high noise on up to time $t = 2$~ns, and then turning the noise off.
The color scale spans from $\rho_1=3.3 \cdot 10^{-4}$ (blue)
to $\rho_1=2.0 \cdot 10^{-3}$ (red).}
\end{figure}

Figure~\ref{fig:turing_hom} illustrates the pattern formation observed in the
system for low noise (top row), high noise (middle row), and deterministic evolution (bottom row) started
from a perturbed initial condition generated by the high noise simulation (see figure caption).
The boundaries take some time to influence the center of the domain,
so in the center the reservoir species are depleted and the system moves toward chemical equilibrium.
However the boundary continuously forces the system so eventually spotted patterns develop,
starting near the boundary, eventually filling the system.
The resulting patterns are qualitatively similar to the ``$\lambda$ pattern''
observed by Pearson~\cite{Pearson09071993} in the GS model.
In simulations with only species diffusion (i.e., setting all other transport to zero) we find
similar patterning, indicating that this system is well-approximated by reaction-diffusion
due to very large solvent concentrations.

\begin{table}
\begin{tabular}{|c||c|c|c|c|c|}
  \hline
  Reaction & $\mathcal{R}_1$ & $\mathcal{R}_2$ & $\mathcal{R}_3$ & $\mathcal{R}_4$ & $\mathcal{R}_5$ \\
  \hline
  $k^+$ & $2.0\cdot10^{-11}$ & $2.0\cdot10^{-11}$ & $3.3333\cdot10^{9}$ & $3.3333\cdot10^{12}$ & $3.3333\cdot10^{12}$ \\
  $k^-$ & $2.0\cdot10^{-17}$ & $2.0\cdot10^{-12}$ & $3.3333\cdot10^{-1}$ & $3.3333\cdot10^5$ & $3.3333\cdot10^5$ \\
  \hline
\end{tabular}
\begin{tabular}{|c||c|c|c|c|c|c|}
  \hline
  Species & $U$ & $V$ & $W$ & $S$ & $U_f$ & $V_f$ \\
  Boundary $n_r$ & $1.513\cdot10^{19}$ & $4.38\cdot10^{18}$ & $3.8\cdot10^{17}$ & $5.0\cdot10^{20}$ & $1.33\cdot10^{20}$ & $5.0\cdot10^{20}$ \\
  \hline
\end{tabular}
\caption{\label{table:BPM_Pattern}
Tables of reference number densities (bottom) and reaction rate parameters (top) for BPM model used to create Fig.~\ref{fig:turing_hom}, in CGS units.}
\end{table}

In \cite{Turing1952} Turing writes, ``Another implicit assumption concerns random disturbing
influences. Strictly speaking one should consider such influences to be continuously at
work. This would make the mathematical treatment considerably more difficult without
substantially altering the conclusions.'' However, we see from Fig.~\ref{fig:turing_hom}
that the evolution is qualitatively different when large spontaneous fluctuations
are present (middle row), as compared to when they are absent (bottom row).
Specifically, the speed at which the patterns develop and propagate is
noticeably accelerated by the spontaneous fluctuations (top and middle row),
though the patterns themselves are qualitatively unchanged in this particular case.
Other studies using reaction-diffusion models and particle simulations have reached similar conclusions~\cite{Turing_Fluctuations1,Turing_Fluctuations2,Turing_RDME,Turing_RDME_2}.
In \cite{Turing_RDME_3} the authors investigated the Gray-Scott model
by RDME simulations and concluded, 'Complex spatiotemporal patterns, including spiral waves, Turing patterns, self-replicating spots and others, which are not captured or correctly predicted by the deterministic reaction-diffusion equations, are induced by internal reaction fluctuations.'

\section{Conclusions and Future Work}
\label{sec:conclusions}

In this work we have formulated a fluctuating hydrodynamics model for chemically reactive
ideal gas mixtures, and developed a numerical algorithm to solve the resulting
system of stochastic partial differential equations. In our Langevin formalism, the stochastic
mass, momentum and heat flux as well as the stochastic chemical production rate, are modeled
using uncorrelated white noise processes, and the local number densities are real variables.
This is contrast to the more traditional
chemical master equation (CME) description of reactions that accounts for every individual reaction
as a small jump of the (potentially very large) integer number of reactant molecules.
We formulated the thermodynamic
driving force for chemical reactions in agreement with nonlinear nonequilibrium
thermodynamics \cite{FluctuatingReactionDiffusion} and considered two different Langevin formulations
of the stochastic chemical production rate. The first formulation is based on the law of mass action
cast in the GENERIC framework \cite{GENERIC_Chemical} and leads to a noise covariance that is the
logarithmic mean (LME) of the forward and reverse production rates. This formulation is fully
consistent with {\em equilibrium} statistical mechanics, more specifically, the resulting
dynamics is time reversible (i.e., satisfies detailed balance)
with respect to the Einstein distribution for a closed system.
The second formulation is based on the
chemical Langevin equation (CLE) \cite{ChemicalLangevin_Gillespie,KeizerBook,KurtzTheorem_CLE},
and while it is not consistent with equilibrium statistical mechanics
this form has its own merits.

We compared the two formulations on two chemical reaction networks
for a well-mixed system, for both a simple dimerization reaction and a more complex
network exhibiting bistability. We confirmed that at thermodynamic equilibrium the LME is
more appropriate than the CLE, however, this is reversed for systems away from equilibrium,
when compared with the CME.
Not unexpectedly, neither is found to be entirely appropriate for describing rare events or large deviations
from equilibrium. These examples remind us that a stochastic differential equation,
which is a diffusion process, cannot be a uniformly accurate approximation
for the CME, which is a Poisson process;
the large-deviation statistics for Poisson noise is different than that of Gaussian noise.
\modified{To further complicate the picture, there are known examples in which the discrete (integer-valued) nature of molecular populations plays a key role, implying that descriptions using real-valued concentrations such as fluctuating hydrodynamics must fail. For example, in wave fronts of the Fisher, Kolmogorov, Petrovski, Piskunov (FKPP) type it has been shown that the discreteness of the ME induces a logarithmic correction to the wave speed~\cite{ChemWaves_MD_ME},
similar to that observed when introducing a small cutoff in the leading edge of a FKPP front~\cite{BrunetDerrida97}
}

Another alternative coarse-grained description of stochastic chemistry, which we did not consider
in this work, is tau leaping \cite{ChemistryReview_Gillespie,WeakTrapezoidal_Chemical,TauLeap_Weak}.
Tau leaping is usually seen as a numerical method for approximately solving the CME,
and the CLE can be seen as an approximation to tau leaping in a specific (central) limit in which
a Poisson and a Gaussian variable become indistinguishable \cite{ChemistryReview_Gillespie};
observe however that the two kinds of distributions always have different tails.
A different and more interesting characterization of tau leaping is to see it instead
as an {\em alternative} to the CLE that maintains the
Poisson nature of the noise rather than replacing it with Gaussian noise.
In the limit that the time step $\Delta t \rightarrow 0$,
for Gaussian noise one gets the CLE, and for Poisson noise one gets, in principle, the CME.
In this limit using the SSA algorithm is an efficient method to solve the CME exactly,
and tau leaping is only useful as a numerical scheme for large $\Delta t$.
As mentioned earlier, computational fluctuating hydrodynamics should
only be considered useful (or even valid) when each update of the coarse-grained degrees of freedom involves an average
over many molecular events, such as many molecular collisions for momentum and energy transport, or many
reactive collisions for chemistry. In other words, to distinguish it from the CME and the associated SSA algorithm,
for tau leaping one should choose the time step size in a way that ensures
that many reactions occur in each reactive channel.
In this sense, tau leaping can be seen as a coarse-graining in time of the CME jump process, and,
when combined with spatial coarse-graining, has the potential to be a useful coarse-grained model
that bypasses the need for Langevin models of chemistry in our numerical schemes for
reactive fluctuating hydrodynamics. Additional studies are needed to access the accuracy of tau leaping
in situations where Langevin descriptions do poorly and such investigations are in progress.

As is well-known, the failure of Langevin approximations to describe large deviations is in fact closely connected to the fact
that traditional linear nonequilibrium thermodynamics fails to describe chemical reactions
because the entropy production rate is generally
a non-quadratic function of the thermodynamic driving force (affinity).
The mesoscopic Kramers picture of chemical reactions,
as developed for isomerization in Ref. \cite{FluctuatingReactionDiffusion}, is an interesting approach, which, however,
remains mostly of theoretical utility; numerical simulations of this model would need to handle additional
dimensions, as well as very slow diffusion across the reaction barrier. Furthermore, it is not
obvious how to extend this description to general multispecies fluids with complex reaction networks.

While the CME is a well-agreed upon and well-justified description of statistically homogeneous systems, the story
is much less clear for systems with spatial inhomogeneity; in fact, the precise mathematical meaning of
``well-mixed'' and the range of validity of the RDME remains obscure.~\cite{MesoRD_GridResolution,RDME_Bimolecular_Petzold,CRDME}
A law of large numbers has been rigorously proven for several particle
models and takes the expected form of a deterministic
reaction-diffusion partial differential equation.~\cite{RD_DeterLimit_Arnold1980,RDME_Limits_Arnold}
Regarding fluctuations, however, there are very few particle models
for which central limit theorems \cite{RD_CLT_Lattice}
or large deviations functionals \cite{RD_Ising_LDT} are known explicitly.
It has been demonstrated that inhomogeneity leads to qualitative changes
in the nature of phase transitions in bistable systems~\cite{SchloglDiffusion_TNL}.
It is also known that fluctuations can effectively renormalize the macroscopic transport
and lead to non-analytic corrections to the law of mass action \cite{Annihilation_AplusB,Coagulation_Renormalization}.
It remains to be seen whether {\em nonlinear} spatially-extended fluctuating hydrodynamics
models can correctly reproduce this effect compared to particle simulations.
\footnote{Fluctuating hydrodynamics simulations have been shown to correctly describe the renormalization of diffusion coefficients in binary fluid mixtures \cite{DiffusionRenormalization}, but we are not aware of any work in this direction for
reactive systems.}

Furthermore, most particle models used for reaction-diffusion problems assume independent
random Brownian walkers that react when coming near each other, and thus completely neglect
transport mechanisms such as advection, sound waves, thermal conduction, etc. Furthermore, the mechanism
of diffusion used in these models implicitly
neglects the long-ranged hydrodynamic correlations
present among particles diffusing in a viscous solvent \cite{DDFT_Hydro}.
Notable exceptions are variants of the Direct Simulation Monte Carlo method (DSMC),
which use a dilute \cite{DSMC_Bird,DSMCReview_Garcia} or dense \cite{SHSD_PRL} gas kinetic theory description
of momentum and energy transport fully consistent
with fluctuating hydrodynamics \cite{SHSD_PRL}. While chemical reactions are commonly
included in DSMC schemes \cite{MacroscopicDSMC_ComplexReactions,MacroscopicDSMC_Reactions},
further studies of fluctuations and their consistency with nonequilibrium thermodynamics
are needed. While some investigations of spatially-distributed reactive systems
have been performed using DSMC \cite{Turing_DSMC}, a careful comparison to coarse-grained mesoscopic
descriptions such as fluctuating hydrodynamics is needed.
To model realistic chemistry modern DSMC codes use either the
Total Collision Energy (TCE) model~\cite{DSMC_Bird}
or the more recent Quantum-Kinetic (QK) model~\cite{BirdQKmodel},
both of which we plan to compare with our formulations of reactive fluctuating hydrodynamics.

In Section~\ref{GiantFluctuationsSection} we studied the coupling of velocity fluctuations and chemistry
in a system kept in a non-equilibrium steady state via boundary conditions.
We found that, in agreement with existing theoretical computations, the chemical reactions have a strong effect on
the giant long-range correlated concentration fluctuations. In this work we focused on gas mixtures,
for which the Schmidt number is of order unity. We found that reactions
profoundly change the nature of the giant fluctuations, whose spectrum
switches from the well-known $\sim k^{-4}$ at large wavenumbers to a flat plateau (with a value controlled
by the reaction rate) at small wavenumbers.
This could be useful, for example, for experimentally measuring reaction rates. However, we found that
the simple quasi-periodic near-equilibrium theory we constructed was not in quantitative agreement with the simulations,
indicating that a more precise theory is needed.

Reactive {\em liquid} mixtures are common in practice and
exhibit interesting coupling of hydrodynamics with transport that has
been studied both theoretically \cite{BuoyancyInstabilities_Chemistry} and experimentally \cite{InstabilityRT_Chemistry}.
The Schmidt number in liquids is, however, very large, and the compressibility is very small (i.e., the speed
of sound is very large), making the method presented here computationally infeasible.
In the future, we will consider
extending low Mach number (quasi-incompressible) methods that
treat momentum diffusion implicitly to include stochastic chemistry, by combining Langevin or tau-leaping
based descriptions of chemistry with the formulation and numerical methods developed in
Refs. \cite{LowMachImplicit,LowMachMultispecies} for general non-ideal liquid mixtures.

Finally, the influence of hydrodynamic fluctuations on reactions
will likely be very important for
surface chemistry.~\cite{CatalysisBook,SurfaceChemistryPnas}
An important application is heterogeneous catalysis in which
a highly reactive catalytic surface facilitates bond breaking
and bond rearrangement of adsorbed molecules.
In this context mesoscale simulations are particularly useful for
the study of nano-catalytic systems~\cite{NanocatalysisReview}.
Microscopic catalytic particles have many advantages, such as
higher activity, increased selectivity, and longer lifetime.
However, to operate effectively these particles must have electrical
contact with a substrate (typically a flat surface) and they must not be
so small as to not have enough electrons to catalyze a reaction.
For example, nanowire catalysts are typically 10-100~nm in size so the
hydrodynamic environment in which the chemistry and transport occur
is of mesoscopic scale (a few microns).

As in the case with chemistry in bulk flow, particle-based simulations
will be useful benchmarks for comparison with reactive fluctuating hydrodynamics
modeling surface chemistry.
Furthermore, molecular simulations
can be embedded within a fluctuating hydrodynamic code to create
an Algorithm Refinement (AR) hybrid.~\cite{Bell_10,FluctuatingHydro_AMAR}
The idea is to use a particle-based simulation for the domain near the
surface to capture the physics at the molecular scale,
such as adsorption of reactants onto the surface,
surface diffusion, surface reactions, and desorption of products.
This particle-based simulation would then be coupled to a fluctuating hydrodynamic simulation
that treats the bulk fluid.
Previous work for non-reactive fluids has already proven the
utility of AR hybrids for simple interfaces~\cite{DSMC_Hybrid}
and this numerical framework should prove useful in the
study of surface reactions and active membrane transport.


\begin{acknowledgments}
We would like to thank M. Malek-Mansour, Jonathan Goodman, Eric Vanden-Eijnden, Samuel Isaacson, Hans Christian Ottinger, Dick Bedeaux, Annie Lemarchand, Florence Baras, John Pearson, Sorin Tanase Nicola and Signe Kjelstrup for informative discussions. This material is based upon work supported by the U.S. Department of Energy Office of Science, Office of Advanced Scientific Computing Research, Applied Mathematics program under Award Number DE-SC0008271 and under contract No. DE-AC02-05CH11231.
\end{acknowledgments}

\begin{appendix}

\section{\label{AppendixHydro}Deterministic and Stochastic Transport in Ideal Gas Mixtures}

This appendix summarizes fluctuating hydrodynamics for ideal gas mixtures; for a more general and detailed
exposition see~\cite{MultispeciesCompressible}.
Each of the hydrodynamic transport terms in (\ref{eq:spec})-(\ref{eq:energy})
contains a deterministic term, denoted with an overbar and a
stochastic term denoted by a tilde (e.g., ${\SpeciesFlux} = \overline{\SpeciesFlux} + \widetilde{\SpeciesFlux}$).
The deterministic viscous tensor is,
\begin{equation}
\overline{\StressTensor} =
-\ShearViscosity \left ( \nabla \mathbf{v} + (\nabla \mathbf{v})^T  \right )
-\left(\BulkViscosity - \frac{2}{3} \ShearViscosity \right) \mathbf{I} \left({\bf \nabla} \cdot {\bf v}\right),
\end{equation}
where $\ShearViscosity$ and $\BulkViscosity$ are the shear and bulk viscosity, respectively.
We neglect any possible effect of
the chemical reactions on the transport coefficients of the mixture.
For example, we neglect any coupling between bulk viscosity and chemical
reactions, which, in principle is allowed by the Curie principle
since both are scalar processes.~\cite{FluctChemistry_Grossmann}
The corresponding stochastic viscous flux tensor can be
written as~\cite{Landau:Fluid,LLNS_Espanol}
\begin{equation}
\widetilde{\StressTensor}(\mathbf{r},t) =
\sqrt{2 k_B T \ShearViscosity}\;
\widehat{\mathcal{Z}}^{\StressTensor}
+ \left ( \sqrt{\frac{k_B \kappa T}{3}}
- \sqrt{\frac{2 k_B \eta T}{3}} \right ) \text{Tr} ( \widehat{\mathcal{Z}}^{\StressTensor}  ).
\label{StressTensorCorrelationEqn}
\end{equation}
The symmetric Gaussian random tensor field, $\widehat{\mathcal{Z}}^{\StressTensor}$,
is formulated as
$\widehat{\mathcal{Z}}^{\StressTensor} = \left( \mathcal{Z}^{\StressTensor} + (\mathcal{Z}^{\StressTensor})^T \right) / \sqrt{2}$
where
$\mathcal{Z}^{\StressTensor}$ is a white-noise random Gaussian tensor field,
that is, $\langle \mathcal{Z}^{\StressTensor} \rangle = 0$ and,
\begin{equation*}
\langle \mathcal{Z}^{\StressTensor}_{\alpha\beta}(\mathbf{r},t)
\mathcal{Z}^{\StressTensor}_{\gamma\delta}(\mathbf{r}',t') \rangle
= \delta_{\alpha\gamma} \delta_{\beta\delta} \,
\delta(\mathbf{r}-\mathbf{r}')\, \delta(t - t') \;\; .
\end{equation*}
with $\alpha,\beta,\gamma,\delta = \{x,y,z\}$ being spatial components.

The deterministic mass flux and heat flux depend on the gradients
of concentration, pressure, and temperature.
For an ideal gas we can write the deterministic fluxes as
\begin{eqnarray}
\SpeciesFlux &=&
- \rho \mathcal{Y} \mathbf{D} \left( \mathbf{d} + \mathcal{X} \tilde{\chi} \frac{\nabla T}{T} \right)
\qquad \mathrm{and}
\label{GasSpeciesFluxEqn}
\\
\HeatFlux  &=&
-\lambda \nabla T + (k_B T \tilde{\chi}^T \mathcal{M}^{-1} + h^T) \SpeciesFlux
\label{GasHeatFluxEqn}
\end{eqnarray}
where $h$ is a vector of specific enthalpies. Here the diffusion driving force \cite{IrrevThermoBook_Kuiken} is
\[
\mathbf{d} = {\nabla} X + \left( X - Y \right) \frac{{ \nabla} p}{p}.
\]
and $\cal{X}$, $\cal{Y}$ and $\cal{M}$
are diagonal matrices of mole fractions $X$, mass fractions $Y$ and molecular masses $M$.
The matrix of multicomponent flux diffusion coefficients, $\mathbf{D}$, the vector of rescaled thermal
diffusion ratios, $\tilde{\chi}$, and the thermal conductivity, $\lambda$, can be obtained from
standard software libraries, such as EGLIB~\cite{EGLIB},
or from standard references, such as Hirshfelder \textit{et al.}~\cite{hirschfelder1967molecular}.

For the ideal gas transport coefficients described above, we define
\begin{equation}
{\mathbf{L}} = \frac{\rho \overline{m}}{k_B} \mathcal{Y} \mathbf{D} \mathcal{Y},
\qquad
{\xi} = k_B T \mathcal{M}^{-1}  \tilde{\chi},
\qquad
\zeta = T^2 {\lambda}.
\end{equation}
where $\overline{m} = (\sum_s Y_s/m_s)^{-1}$ is the mixture-averaged molecular mass.
The stochastic terms for the combined species equations and energy equation are determined from
the phenomenological equations of nonequilibrium thermodynamics that relate fluxes
to thermodynamic driving forces through the Onsager matrix,
and the fluctuation-dissipation balance principle.~\cite{MultispeciesCompressible}
Specifically, the stochastic mass flux is
\begin{equation}
\widetilde{\SpeciesFlux}  = \mathbf{B} \; \ZSpeciesFlux
\label{MassDiffNoiseEqn}
\end{equation}
where $\mathbf{B} \mathbf{B}^T = 2 k_b \mathbf{L}$, and
$\ZSpeciesFlux$ is a white-noise random Gaussian vector field with uncorrelated components,
that is, $\langle \ZSpeciesFlux \rangle = 0$ and
\[
\langle \ZSpeciesFluxC_{\alpha s}(\mathbf{r},t)
\ZSpeciesFluxC_{\beta s'}(\mathbf{r}',t') \rangle
= \delta_{\alpha\beta} \delta_{s,s'} \,
\delta(\mathbf{r}-\mathbf{r}')\, \delta(t - t')
\]
with $\alpha,\beta = \{x,y,z\}$ being spatial components.
Note that the matrix $\mathbf{B}$ can be obtained using Cholesky decomposition;
the constraint $\sum_s \widetilde{\SpeciesFlux}_s = 0$ is ensured by construction.~\cite{MultispeciesCompressible}
The stochastic heat flux is
\begin{equation}
\widetilde{\HeatFlux} =
\sqrt{\zeta} \ZHeatFlux +
(\xi^T + h^T) \widetilde{\SpeciesFlux},
\label{HeatDiffNoiseEqn}
\end{equation}
where
$
\langle \ZHeatFluxC_{\alpha}(\mathbf{r},t)
\ZHeatFluxC_{\beta}(\mathbf{r}',t') \rangle
= \delta_{\alpha\beta} \, \delta(\mathbf{r}-\mathbf{r}')\, \delta(t - t')
$.

\section{\label{AppendixDimerization}Equilibrium distribution for a dimerization reaction}

The entropy of mixing for a system undergoing dimerization is $S=k_B\ln N_p$, where $N_p$ is the
number of distinct ways of forming $M$ dimers out
of a total of $N$ monomers. This is straightforward to compute.
The number of ways to choose $2M$ out of the $N$ atoms to be in pairs is,
\[
\left(\begin{array}{c}
N\\
2M
\end{array}\right)=\frac{N!}{\left(2M\right)!\left(N-2M\right)!}
\]
Next we need to group the $2M$ atoms into pairs; the
number of distinct ways of pairing $2M$ objects is
$
{(2M)!}/{2^{M}M!}.
$
This gives the entropy
\begin{equation}
S = k_{B}\ln\left(\frac{N!}{2^{M}M!\left(N-2M\right)!}\right) + k_B \mu M,
\label{eq:counting}
\end{equation}
where we have included an additional reference chemical potential $\mu$ to
set the equilibrium concentration.

By expanding the logarithm of the right-hand side of (\ref{eq:counting})
using Stirling's leading-order approximation we obtain the thermodynamic
limit of the entropy as a function of the monomer mass fraction $Y_1=\left(N-2M\right)/N$.
After fixing the chemical potential from the requirement that the
most probable mass fraction is
$(Y_1)_\text{eq}=1/2$, we get,
\begin{equation}
S=Nk_{B}\left[-\frac{1}{2}\,\ln\left(1-Y_1\right)\left(1-Y_1\right)
-Y_1\ln\left(Y_1\right)-\frac{1}{2}\, Y_1\left(\ln\left(2\right)-1\right)\right],
\label{eq:entropy_counting}
\end{equation}
This gives an Einstein distribution $P \sim e^{S/k_{B}}$ exactly matching
(\ref{eq:P_eq_Hanggi}).

It is useful to compare this equilibrium distribution of the LME with
that of the CME for a well-mixed system of volume $\D V$. It is not hard to show
that the CME for a dimerization reaction is in detailed balance with respect to the Einstein
distribution with entropy (\ref{eq:counting}), with the reference chemical potential set to
\begin{equation}
\mu=\ln\left(\frac{2k^{+}}{\D V\, k^{-}}\right).
\label{eq:mu_AA}
\end{equation}
We can use this to construct a rather accurate continuum approximation to this exact microscopic
result by including the next order term in the Stirling formula,
\begin{equation}
N!\approx\sqrt{2\pi N}\left(\frac{N}{e}\right)^{N},
\label{eq:Stirling}
\end{equation}
when expanding (\ref{eq:counting}).
This gives a better continuous approximation (labeled ``Stirling'' in the figures in the main body of this paper)
to the discrete Einstein distribution than (\ref{eq:entropy_counting}),
especially for small number of particles in the well-mixed cell.

\section{\label{AppendixGiant}Giant fluctuations in the presence of reactions}

This appendix outlines the fluctuating hydrodynamics theory for
the long-range correlations of concentration fluctuations in a binary mixture undergoing a dimerization reaction,
as studied numerically in Section \ref{GiantFluctuationsSection}.
We neglect the Dufour effect and assume the system to be isothermal,
taking contributions from temperature fluctuations to be of higher order.
Furthermore, we neglect gravity, assume the system is incompressible,
and take the density and transport coefficients to be constant.
We consider a ``bulk'' system \cite{FluctHydroNonEq_Book},
i.e., we neglect the influence of the boundaries.
This gives an accurate approximation for wavenumbers that are large compared
to the inverse height of the domain; for smaller wavenumbers the
boundaries are expected to suppress the giant fluctuations~\cite{FluctHydroNonEq_Book,GiantFluctFiniteEffects}.
We also neglect the stochastic mass flux in the concentration equation
since we are concerned with the nonequilibrium contribution due to the forcing by the velocity fluctuations.

We assume that all of the concentration gradients are in the same
direction (say, the $y$ axis). The incompressibility constraint
is most easily handled by applying a $\grad\times\grad\times$
operator to the momentum equation ~\cite{FluctHydroNonEq_Book} to obtain a system involving
only the component of the velocity parallel to the gradient, $v_{\parallel} \equiv v_y$.
The same calculation can easily be generalized to a multispecies mixture as well,
see Appendix B in Ref. \cite{MultispeciesCompressible}.
This system of equations can be most easily solved in the Fourier
domain, by considering wavevectors $\V{k}$ in the plane
perpendicular to the gradient, $\V{k} = \V k_{\perp}$.

It is very straightforward to derive (\ref{eq:S_AA}) in the limit of large Schmidt
number by considering the fluctuating concentration equation forced
by an \emph{overdamped} (steady Stokes) fluctuating velocity. The
steady Stokes equation in Fourier space has the form,
\[
\eta k^{2}\hat{v}_{\parallel}=\sqrt{2\eta k_{B}T_0}\; ik\widehat{\mathcal{W}}(t),
\]
where $T_0$ is the constant temperature.
Here $\widehat{\mathcal{W}}(t)$ is a white-noise process (one per wavenumber),
giving the white-in-time velocity $\hat{v}_{\parallel}(t)=k^{-1}\sqrt{2k_{B}T_0/\eta}\; i\widehat{\mathcal{W}}(t)$.
A general form of the linearized equation for
the concentration fluctuations, $c = \delta Y_1 = Y_1 - \av{Y_1}$,
in Fourier space will have the form
\[
\partial_{t}\hat{c}=- D  k^{2}\hat{c}-\psi\hat{c} - \hat{v}_{\parallel} f
=-\left( D  k^{2}+\psi\right)\hat{c} - i f \sqrt{\frac{2k_{B}T_0}{\eta k^{2}}}\widehat{\mathcal{W}}(t),
\]
where $\psi$ is the reaction rate at the equilibrium point (around which
we are linearizing), and $f=d\av{Y_1}/dy$ is the applied concentration gradient.
For our specific reaction and the choice of equilibrium point
$(Y_1)_{\text{eq}}=1/2$, we have that $\psi=3k^{-}$ (to see this simply linearize (\ref{eq:DimerizationRD}) around $(Y_1)_{\text{eq}}=1/2$).
The resulting nonequilibrium static structure factor
(recall that here we neglect the contribution due to the stochastic mass flux) is straightforward to calculate
(see, for example, Appendix B in Ref. \cite{MultispeciesCompressible}),
\begin{equation}
S_{\text{neq}}(k)=\av{\hat{c}\hat{c}^{\star}}=\frac{k_{B}T_0}{\eta k^{2}\left( D  k^{2}+\psi\right)}f^{2}=\frac{k_{B}T_0}{\eta D  k^{4}\left(1+\psi D ^{-1}k^{-2}\right)}f^{2},
\label{eq:S_AA_simple}
\end{equation}
which is exactly (\ref{eq:S_AA}) with the penetration depth
\[
d^2=\frac{D}{\psi}=\frac{ D }{3k^{-}}.
\]

Since the Schmidt number is not very large for gases, we should improve
the theory by not taking the overdamped limit but rather adding a
velocity equation and considering the linearized {\em inertial} equations,
\begin{eqnarray*}
\partial_{t}\hat{c} & = & - D  k^{2}\hat{c}-\psi\hat{c} - \hat{v}_{\parallel} f \\
\rho_0 \partial_{t}\hat{v}_{\parallel} & = & -\eta k^{2}\hat{v}_{\parallel}+\sqrt{2\eta k_{B}T_0}\; ik\widehat{\mathcal{W}}(t),
\end{eqnarray*}
where $\rho_0$ is the equilibrium density and $\nu = \eta/\rho_0$ is the kinematic viscosity.
Solving this system of linear SODEs gives the improved structure factor
\begin{equation}
S_{\text{neq}}(k)=\frac{k_{B}T_0}{\eta D  k^{4}\left(1+\psi D ^{-1}k^{-2}\right)\left(\frac{ D +\nu}{\nu}+\psi\nu^{-1}k^{-2}\right)}f^{2},
\label{eq:S_AA_rnertial}
\end{equation}
which shows that the finite value of the Schmidt number $S_{c}=\nu/D$ has an important effect on the giant fluctuations. In particular, in the more complete theory (\ref{eq:S_AA_rnertial})
\modified{
\[
S_{\text{neq}}(k=0)=\frac{k_{B}T_0}{\rho_0 \psi^2}f^{2}
\]
}
is finite and not infinite as in (\ref{eq:S_AA_simple}), which assumes infinite Schmidt number.

In the more complete theory we see the following three regimes:
\begin{enumerate}
\item For large wavenumbers we have
\[
S_{\text{neq}} \left( k\gg\sqrt{\frac{ D }{\psi}} \right) \approx
\frac{k_{B}T_0}{\eta D  k^{4}}f^{2}
\]
as if there were no reaction.
\item If the Schmidt number is small, $ D \sim\nu$, as for gases, then
for small wavenumbers we instantly switch to a flat spectrum
\[
S_{\text{neq}} \left( k\ll\sqrt{\frac{ D }{\psi}} \right) \approx\frac{k_{B}T_0}{\rho_0 \psi^2}f^{2}
\]

\item If the Schmidt number $S_{c}$ is large, $ D \ll\nu$, as for liquids,
then for intermediate wavenumbers
\[
S_{\text{neq}} \left(  \sqrt{\frac{ D }{\psi}}\ll k\ll S_{c}^{\frac{1}{2}}\sqrt{\frac{ D }{\psi}} \right) \approx
\frac{k_{B}T_0}{\eta \psi  k^{2}}f^{2}
\]
we observe a change in the power law to $S_{\text{neq}}(k)\sim k^{-2}$.
Note however that this range spans only $\sqrt{S_{c}}$ orders of
magnitude in $k$, so even for $S_{c}\sim10^{4}$ (which applies to macromolecular solutions),
the $k^{-2}$ power-law only extends over at most two decades. Therefore, even for liquids with
large Schmidt numbers the inertial equations should be used to model giant fluctuations
in reactive mixtures.
\end{enumerate}

\end{appendix}

%

\end{document}